\documentclass[trackchanges]{aastex7}
\usepackage{amsmath}
\usepackage{graphicx}
\usepackage{subcaption}
\usepackage{multirow}

\begin{document}
\title{Measuring Outflow Distances in NGC 5548 Using Absorption-Line Variability Diagnostics}

\author{Yaqi Chen}
\email{}
\affiliation{Department of Astronomy, University of Science and Technology of China, Hefei, Anhui 230026, China
}%
\affiliation{School of Astronomy and Space Science, University of Science and Technology of China, Hefei, Anhui 230026, China
}

\correspondingauthor{Zhicheng He}
\email{zcho@ustc.edu.cn}

\author{Zhicheng He}
\email{}
\affiliation{Department of Astronomy, University of Science and Technology of China, 
 Hefei, Anhui 230026, China
}%
\affiliation{School of Astronomy and Space Science, University of Science and Technology of China, Hefei, Anhui 230026, China
}%
 %\altaffiliation{}%Lines break automatically or can be forced with \\

 \correspondingauthor{Guilin Liu}
\email{glliu@ustc.edu.cn}
\author{Guilin Liu}%
\email{}
\affiliation{Department of Astronomy, University of Science and Technology of China, 
 Hefei, Anhui 230026, China
}%
\affiliation{School of Astronomy and Space Science, University of Science and Technology of China, Hefei, Anhui 230026, China
}%

\begin{abstract}

AGN-driven outflows serve as a key channel through which the energetic central engine influences host galaxy evolution. Among the physical properties of outflows, their radial distance from the galactic nucleus is particularly important for assessing AGN feedback. In this study, we investigate the UV outflow components in NGC 5548 by analyzing the variability of C IV absorption troughs in optical spectra obtained through multiple HST observations during 2013 and 2014. We construct a set of variability-based diagnostic events, labeled G1 and G2, which are sensitive to the recombination timescale ($t_r$) of ionized gas. By combining these with mock light curves generated using a damped random walk (DRW) model, we numerically establish a mapping between the G1 event probability and $t_r$. This approach allows us to constrain the radial distances of outflow components 1 and 6, whose absorption variability is primarily driven by changes in the incident ionizing continuum, to be $0.77^{+0.10}_{-0.10}$ and $1.72^{+1.74}_{-1.72}$ pc, respectively. These results are consistent with those obtained using a different method in our previous study, as well as with values reported in the literature.

\end{abstract}
% AGN驱动的外流是中心高能引擎影响宿主星系活动的重要桥梁，其中，外流到星系中心的距离是研究AGN反馈的重要参数。我们使用2013和2014年间HST多次观测的光学光谱通过 C IV 吸收坑光变来研究NGC 5548的 UV 外流成分的半径距离。我们通过构造与电离气体的复合时标 $t_r$ 相关的光变事件G1/G2作为探针，结合DRW模型生成模拟的光曲线进行数值模拟来获得$t_r$与G1事件概率之间的映射关系，由此测出了吸收线光变是入射电离连续谱造成的外流成分1和6的距离，分别为$0.77^{+0.10}_{-0.10}$ and $1.72^{+1.74}_{-1.72}$ pc。该结果与其它方法和其它文献的测量结果相近。

%\keywords{\uat{Galaxies}{573} --- \uat{Cosmology}{343} --- \uat{High Energy astrophysics}{739} --- \uat{Interstellar medium}{847} --- \uat{Stellar astronomy}{1583} --- \uat{Solar physics}{1476}}

\section{Introduction}
Supermassive black holes (SMBHs) are commonly found at the centers of galaxies, and they grow primarily through episodes of gas accretion. During these phases of rapid accretion, SMBHs are observed as active galactic nuclei (AGNs) (\cite{1969Natur.223..690L,1984ARA&A..22..471R}). 
A hallmark of AGNs is their enormous electromagnetic output, which spans a wide range of wavelengths—from the infrared (IR) to gamma rays. The accretion process liberates tremendous energy, which powers winds and/or relativistic jets. These energetic outflows can expel or heat the interstellar gas in the host galaxy, thereby suppressing or triggering star formation. This so-called AGN feedback is considered a key mechanism in the co-evolution of SMBHs and their host galaxies. (\cite{2005MNRAS.361..776S,2006ApJS..163....1H,2006MNRAS.370..645B,2007ApJ...665..120A,2008MNRAS.390.1399B,2008MNRAS.391..481S,2010MNRAS.406..822M,2022NatAs...6..339C,2024SCPMA..6729512H})

The strength of AGN feedback can be quantitatively assessed by characterizing the physical properties of the outflows, such as the mass of the outflow, mass outflow rate, and kinetic luminosity (\cite{2012ApJ...751..107B}).
Among these, key parameters include the outflow velocity, the ionic column density, and the distance from the central AGN. The first two quantities can typically be inferred from the width of absorption lines in spectra combined with photoionization modeling. However, the most significant source of uncertainty often lies in determining the outflow’s distance from the nucleus.

For nearby galaxies, the spatial extent of AGN-driven outflows can be directly resolved and measured using integral field unit (IFU) spectroscopy. In most other cases, however, the distance of the outflow is inferred indirectly through the analysis of absorption lines in the spectrum, combined with the definition of the ionization parameter. Since this parameter depends on both the gas density and its distance from the ionizing source, the problem effectively reduces to constraining the density of the outflowing gas (e.g. \cite{2014MNRAS.444.1893C, 2014ApJ...783...58L, 2011ApJ...728...94L, Arav_2015, Arav_Liu_2018, 2018ApJ...866....7L, 2019MNRAS.483.1808H, Ebrero_2016,2017A&A...607A.100M}).
A commonly used method involves examining absorption features from ions in excited states. By comparing the column densities of excited and ground (resonant) levels, the number density can be inferred. However, this approach places stringent demands on spectral quality, requiring multiple, unblended transitions with sufficient signal-to-noise. In AGNs, where the outflows are typically fast and broad, such lines are often limited. %Additionally, in highly luminous sources, intense ionizing radiation can fully strip electrons from many ions, making them undetectable.
High-velocity outflows are generally believed to exert stronger feedback on their host galaxies, and are therefore more relevant for understanding AGN-galaxy co-evolution. Ironically, these are also the regimes where traditional diagnostics become least effective. This disconnect between the limitations of established methods and the focus of current AGN research underscores the need for alternative observational strategies.

In recent years, an alternative approach has emerged for estimating outflow distances (\cite{He_2019, He_2022, 2025A&A...703A.305H,2021ApJ...906L...8Z}), which leverages temporal monitoring of absorption-line variability . This method is based on constraining the recombination timescale $t_r$ (\cite{Arav_2012}), defined as the characteristic delay with which the absorbing gas responds to changes in the ionizing continuum. Two distinct techniques have been proposed to infer $t_r$, and consequently determine the outflow radius.
The first method (\cite{He_2019}) analyzes the variation detection probability as a function of time separation between observations. A rapid increase in detection probability is expected near the recombination timescale. By fitting this probability distribution with a Gaussian cumulative distribution function (CDF), the value of $t_r$ can be constrained.
The second method (\cite{He_2022}) focuses on constructing variability events specifically designed to be sensitive to $t_r$. Through simulations of AGN continuum variability using the Damped Random Walk (DRW) model (\cite{DRW_2009, DRW_2010}), a quantitative relation is established between the occurrence rates of such events and the recombination timescale, enabling an indirect measurement of $t_r$.

Although both of these methods were initially applied to large statistical samples of quasars, involving hundreds of sources, they are fundamentally statistical techniques in nature and based on physical principles that are equally applicable to individual objects. The underlying physical assumptions, such as the delayed response of ionized gas to continuum variations and the modeling of AGN light curves as damped random walks (DRW), are equally valid for individual sources. Therefore, these techniques can also be used to estimate the outflow distance in single-object cases.
Indeed, in addition to presenting results based on large quasar samples, \cite{He_2022} also demonstrated a successful application of the method to the individual AGN SDSS J141955.26+522741.1.
With the continued expansion of time-domain surveys and the growing availability of multi-epoch spectroscopic observations for individual AGNs, the applicability of absorption-line variability techniques to measuring outflow distances is expected to increase steadily in the coming years.

NGC 5548 is a nearby, prototypical Seyfert 1 galaxy and has long served as a key target for AGN studies. As one of the most extensively monitored nearby AGNs, its fundamental physical properties, including variability behavior, spectral energy distribution (SED), and black hole mass, broad-line region (BLR) size, have been thoroughly investigated (\cite{2000A&A...354L..83K, 2013A&A...551L...6K, 2016A&A...592A..27C, 2017MNRAS.464.1771S, 2020ApJ...902...74W, 2022MNRAS.516.4397K, AGN_STORM_2021, Mehdipour_2024}). It remains one of the best-characterized AGNs to date.
The wealth of accumulated spectroscopic observations of NGC 5548 makes it particularly well-suited for detailed studies. In particular, the 2014 AGN Space Telescope and Optical Reverberation Mapping (AGN STORM, \cite{AGN_STORM}) campaign provided an exceptionally high-cadence dataset, with daily monitoring over a six-month period using Swift, the Hubble Space Telescope (HST), and numerous ground-based optical facilities. This intensive monitoring makes NGC 5548 an ideal candidate for estimating outflow distances through absorption-line variability analysis.

This work represents the second study aimed at determining the outflow distance in NGC 5548  based on the variability of C IV absorption lines in the optical spectrum. In the previous work (\cite{Chen_2026}), the distances of six outflow components were estimated using detection probability curves constructed from the variability behavior of absorption troughs labeled A to I. However, for components 1 and 4, only upper limits on the outflow distance were obtained. To further refine the distance measurements, the present study adopts the same methodology proposed in \cite{He_2022}, which involves constructing G1 and G2 events associated with recombination timescales.
Among the six outflow components, only components 1, 5, and 6 exhibit clean, unblended absorption troughs suitable for this type of analysis. Compared to the detection probability curve method, this technique relies more heavily on the assumption that the variability of absorption lines is driven by changes in the incident ionizing continuum. Consequently, component 5 is excluded from the analysis due to its previously observed lack of correlation between continuum flux at 1500 Å and the equivalent width (EW) of the absorption trough (see Figure 5 in the earlier paper, \citet{Chen_2026}).
This study proceeds to identify G1 and G2 events for the remaining components 1 and 6, estimate the corresponding recombination timescales, and ultimately derive the outflow distances for these components.

% 之前该方法仅被用于过高红移的宽吸收线类星体中，这次应用于被大量研究的著名的低红移源NGC5548，也有定量检验该方法的可靠性的考虑在。特别是成分1的结果，它曾被通过传统的激发态离子柱密度方法可靠地测量过外流距离(\cite{Arav_2015})。本文使用G1事件方法更精确地估计出它的半径距离，并与\citet{Arav_2015}（特别是使用啦相同的Obscured SED时）以及之前的探测率方法的结果符合得很好。
%Previously, this method has mainly been applied to high-redshift broad absorption line quasars. Its application here to the well-studied, nearby AGN NGC 5548 also provides an opportunity to quantitatively assess the reliability of the technique. This is particularly relevant for component 1, whose outflow distance has been robustly measured using the traditional method based on excited-state ionic column densities (\cite{Arav_2015}). Using the G1 event approach, we obtain a more precise estimate of its radial distance.%%Using the G1 event approach, we obtain a more precise estimate of its radial distance, which is in good agreement with the results of \citet{Arav_2015}, especially when adopting the same obscured SED, as well as with those derived from the detection probability curve method in our previous work (see Figure~\ref{fig:R_comparison}).
Previously, this method has mainly been applied to high-redshift broad absorption line quasars. Its application here to the well-studied, nearby AGN NGC 5548 also provides an opportunity to quantitatively assess the reliability of the technique, particularly for component 1, whose outflow distance has been robustly measured using the traditional method based on excited-state ionic column densities (\cite{Arav_2015}).

\section{Data}
\label{sec:2}
To investigate absorption-line variability in NGC 5548, we utilize a set of multi-epoch archival ultraviolet (UV) spectroscopic observations. These multi-epoch data are essential for tracing temporal changes in absorption features and diagnosing the physical conditions of AGN outflows. The core dataset consists of spectra from the 2014 AGN Space Telescope and Optical Reverberation Mapping (AGN STORM) campaign, supplemented by earlier observations obtained in 2013 with the Cosmic Origins Spectrograph (COS) on board the Hubble Space Telescope (HST).

The AGN STORM project carried out an intensive six-month monitoring program of NGC 5548, spanning from February 1 to July 27, 2014. Observations were performed on a near-daily cadence using single-orbit exposures with HST/COS, resulting in high-cadence UV coverage. Each visit employed the G130M and G160M gratings with multiple central wavelengths and focal-plane positions (FP-POS), covering the 1153–1796 Å range. Specifically, each visit included two exposures with the G130M grating (centered at 1291 Å and 1327 Å, 200 seconds each) and two with G160M (centered at 1600 Å and 1623 Å, 590 seconds each). This configuration was optimized to achieve a signal-to-noise ratio (S/N) exceeding 100 per $\sim \ 500 km\ s^{-1}$ velocity bin, allowing for reliable detection of subtle absorption variability.

All spectra were retrieved from the Mikulski Archive for Space Telescopes (MAST)\footnote{https://mast.stsci.edu/portal/Mashup/Clients/Mast/Portal.html} and processed with the standard CalCOS pipeline. The relative flux calibration uncertainty is within 2\%.

A number of subsequent studies (\cite{2016ApJ...824...11G, 2019MNRAS.486.5362G,2017ApJ...837..131P}) identified an anomalous phase during the STORM campaign, from JD 2456766.1 to JD 2456829.8 (approximately 60 days), during which the UV–optical broad emission lines became decoupled from the far-UV continuum. This behavior has been attributed either to changes in the unobservable ionizing continuum—affecting its strength, spectral shape, or temporal structure—or to a temporary obscuring structure located between the ionizing source and BLR, which selectively attenuates the ionizing flux reaching the BLR without altering the continuum observed along our line of sight.

Since our analysis assumes that absorption-line variability reflects changes in the incident ionizing radiation, data from the anomalous period are excluded. We therefore restrict our study to epochs prior to JD 2456766.1. (For reference, this corresponds to the unshaded region preceding the gray-shaded anomaly shown in Figure 8 of \cite{2019ApJ...881..153K}.)

To extend the temporal baseline, we include additional UV spectra obtained in 2013 as part of earlier HST/COS monitoring campaigns (\cite{Kaastra_2014_ngc5548_2013_campaign, Mehdipour_2015_ngc5548_2013_campaign}). This set comprises five visits between June 22 and August 1, 2013, each consisting of two HST orbits using both G130M and G160M gratings with multiple central wavelength and FP-POS settings. These observations cover the 1132–1801 Å range at a resolving power of $\sim$15,000.

In total, our final dataset consists of 76 high-resolution UV spectroscopic observations obtained over the 2013 and 2014 epochs.

\section{Methodology and Analysis}
\label{sec:3}

To estimate the radial distance of AGN outflows using absorption-line variability, we employ a method that relies on two fundamental relations, as outlined in \cite{Arav_2012}.

% 假设入射电离连续谱变化前，外流气体正处于电离复合平衡。某一元素在电离阶段 i 的丰度可以写成 $\frac{dn_i}{dt}=-n_i\left(I_i+R_{i-1}\right)+n_{i-1}I_{i-1}+n_{i+1}R_i$。其中 $I_i$ 为单位粒子的电离率；$R_i=\alpha_i(T)\,n_{\mathrm{e}}$ 为从电离阶段 i+1 复合到 i 的单位粒子复合率。通过一系列从电离阶段i到0的联立，层层代入后，得到 $n_iI_i=n_{i+1}R_i$ 的关系式。而某一时刻入射电离连续谱突然发生变化，表述为 $I_i(t>0)=(1+f)\,I_i(t=0)$，其中 $-1\le f\le\infty$。此时，令 $dn_i/dt\rightarrow n_i/t$，再将各种关系式代入，则大致得出复合时标的表达式：
Following \cite{Arav_2012}, the outflowing gas is assumed to be in photoionization equilibrium prior to a change in the incident ionizing continuum. The abundance of ionization stage i for a given element can then be expressed as $\frac{dn_i}{dt}=-n_i\left(I_i+R_{i-1}\right)+n_{i-1}I_{i-1}+n_{i+1}R_i$,
where $I_i$ is the ionization rate per particle and $R_i=\alpha_i(T)\,n_{\mathrm{e}}$ is the recombination rate per particle from ionization stage i+1 to i. By successively combining the coupled equations for all ionization stages from i to 0, one obtains the equilibrium relation $n_iI_i=n_{i+1}R_i$.
A sudden change in the incident ionizing continuum is then parameterized as $I_i(t>0)=(1+f)\,I_i(t=0)$, where $-1\le f\le\infty$. 
Approximating $dn_i/dt$ as $n_i/t$ yields an approximate expression for the recombination timescale (see Section 4 of \cite{Arav_2012} for the derivation):

\begin{equation}
t_r = \left[-f\alpha_{\mathrm{CIV}}n_e\left(\frac{n_{\mathrm{CV}}}{n_{\mathrm{CIV}}} - \frac{\alpha_{\mathrm{CIII}}}{\alpha_{\mathrm{CIV}}}\right)\right]^{-1},  \label{eq:1}
\end{equation}

Here, $t_r$ denotes the recombination timescale of the absorbing gas. The parameter $f$ represents the fractional change in the ionizing continuum between two epochs, for which we adopt a typical value of 0.1, following \cite{He_2022}. The recombination coefficient $\alpha_i$ refers to the recombination rate coefficient from ionization stage i+1 to i, with values obtained from the CHIANTI atomic database (version 8.0; \cite{CHIANTI}) at a temperature of $2 \times 10^4\,\mathrm{K}$. This yields $\alpha_{\mathrm{C\,III}} = 2.1 \times 10^{-11}\,\mathrm{cm^3\,s^{-1}}$ and $\alpha_{\mathrm{C\,IV}} = 5.3 \times 10^{-12}\,\mathrm{cm^3\,s^{-1}}$.

%The ion density ratio $n_{\mathrm{CV}}/n_{\mathrm{CIV}}$ is obtained from photoionization equilibrium models generated using the CLOUDY code \citep{Cloudy2023}, based on the best-fit ionization parameter $U_H$ and total hydrogen column density $N_H$ (see Appendix A of our previous paper, in prep., for model details).
The ion density ratio $n_{\mathrm{CV}}/n_{\mathrm{CIV}}$ is obtained from photoionization equilibrium models generated using the CLOUDY code \citep{Cloudy2023}, based on the best-fit ionization parameter $U_H$ and total hydrogen column density $N_H$. All photoionization parameters, including $U_H$, $N_H$, and the adopted spectral energy distribution (SED), are identical to those used in our previous study (\cite{Chen_2026}). Detailed modeling procedures and parameter values can be found in Appendix A of that paper.

We emphasize that Equation~(\ref{eq:1}) is derived under the assumption of ionization–recombination equilibrium, meaning that $t_r$ characterizes the combined response of the gas to both ionization and recombination processes. %This timescale is therefore best defined in regimes where the flux variation is relatively small.

\begin{equation}
U_H = \frac{Q_H}{4 \pi R^2 n_H c}, \label{eq:2}
\end{equation}

Equation~(\ref{eq:2}) provides the classical definition of the ionization parameter $U_H$, where $Q_{\rm H}$ is the emission rate of hydrogen-ionizing photons (i.e., those with energies above 13.6 eV), R is the distance between the absorber and the ionizing source, $n_H$ is the total hydrogen number density, and c is the speed of light. Given that C IV traces highly ionized gas, we adopt the approximation $n_H \approx 0.83\,n_e$, assuming a fully ionized medium dominated by hydrogen and helium, with negligible contributions from metals.

In this framework, variability in absorption lines is used to constrain the recombination timescale $t_r$, which in turn allows us to estimate the absorber’s radial distance from the central ionizing source when combined with photoionization modeling.

%---------------------

\subsection{Detection of Absorption Line Variability}
\label{sec:3.1}

To identify variability in absorption troughs associated with AGN outflows, we adopt a direct spectral comparison technique instead of relying on traditional EW measurements derived from continuum fitting. 
% 具体来讲，我们将不同时间点的两条光谱的非吸收部分（即连续谱水平以及发射线部分）进行macth，随后一一pixel比较来寻找光变区间。具体步骤可见该系列的前作 \cite{Chen_2026}。
\textbf{Specifically, the unabsorbed portions of two spectra obtained at different epochs, including the continuum and emission-line components, are first matched. The spectra are then compared on a pixel-by-pixel basis to identify variable absorption regions. A detailed description of the procedure is provided in our previous paper in this series (\cite{Chen_2026}).}

The integrated significance of each variable region is computed as:
$$N_\sigma = \frac{|\sum \Delta \mathrm{flux}|}{\sqrt{\sum \sigma^2}}$$.
Only regions with $N_\sigma > 3$ are considered statistically significant. %Representative examples of detected variability are shown in Figures~\ref{fig:var_examples_short} and \ref{fig:var_examples_long}.
% 光谱变化以及匹配和光变的例子可见 \cite{Chen_2026} 的图10和图11
Examples of the spectral variability, as well as the matching and variability classifications, are shown in Figures 10 and 11 of \cite{Chen_2026}.

\subsection{Measuring the Recombination Timescale $t_r$ via G1 Events Probability}

%In cases where the observational cadence is sparse and uneven, or when most time intervals are longer than the recombination timescale $t_r$, the detection rate curve method becomes less effective. Moreover, even within the same object, different velocity or ionization components of the outflow may have significantly different $t_r$ values, and some components may respond on timescales much shorter than the observational cadence. In such cases, their variability may not be effectively captured by detection rate statistics.

%To address this limitation, we employed a probabilistic method based on “G1/G2 events” to constrain $t_r$. 

\subsubsection{Constructing G1 Probability as a Probe of $t_r$}
%Principles Behind the Experimental Design
%Construction of G1/G2 events
%Constructing the G1 Probability as a Probe of $t_r$
%G1 Probability Construction for Probing $t_r$
%我们的目的是要通过光变来确定吸收气体的复合时标，因此需要一个与复合时标相关的观测量，通过这个观测量来约束复合时标。下面将要介绍的“G1事件的观测概率”就是这样一个观测指标。
%我们可以想象下面这个场景：当某个源有任意三次观测时，总能取出一对观测间隔较短的光谱和另一对观测间隔相对较长的光谱，且这两对光谱之间有一个光谱是共有的。当较短时间间隔的光谱对之间有更强的入射电离连续谱变化时，我们就能发现，这两对光谱的光变情况就与复合时标 $t_r$ 的大小（相对于观测间隔的大小）息息相关。
%假设两对光谱的观测时间间隔分别为 $\Delta T_\mathrm{short}$ 和 $\Delta T_\mathrm{long}$，当 $ t_r < \Delta T_\mathrm{short} < \Delta T_\mathrm{long}$ 时，较短间隔的光谱因为有更强的入射光光变以及足够多的时间让吸收气体反应入射光的变化，于是短间隔的光谱就会展现出比长间隔光谱更强的吸收坑光变；当$\Delta T_\mathrm{short} < t_r < \Delta T_\mathrm{long}$ 时，虽然短间隔的光谱有更强的入射光变化，但吸收气体没有足够的时间响应，就会出现长间隔光谱的吸收坑光变大于短时间间隔的情况，出现概率随着 $t_r$ 远离 $\Delta T_\mathrm{short}$ 的程度而逐渐加大；当 $\Delta T_\mathrm{short} < \Delta T_\mathrm{long} < t_r$ 时，吸收线的光变情况就会陷入“混沌”状态，其相对大小取决于入射电离连续谱的变化情况和吸收气体的响应方式，随着$t_r$ 远离 $\Delta T_\mathrm{long}$，两对光谱的吸收线光变应该会趋于相近，因此，如果观测光谱中的最大观测间隔小于$t_r$时，该方法就不适用了。
%因此，在三次观测之间较短时间间隔的光谱对之间有更强的入射电离连续谱变化的前提下，我们将“短时间间隔光谱的吸收线光变大于长时间间隔光谱的吸收线光变”的情况命名为“G1”，“长时间间隔光谱的吸收线光变大于短时间间隔光谱的吸收线光变”的情况命名为“G2”。通过上面的分析可知，“G1事件的发生概率”是一个随着复合时标 $t_r$增长而单调递减的物理量，可用于在观测上约束$t_r$的大小。

Our goal is to constrain the recombination timescale of the absorbing gas by analyzing absorption-line variability. To this end, we require an observable that is directly related to the recombination timescale, serving as a probe for it. The “probability of observing G1 events” introduced below is such an observable.

Consider the following scenario: for any source with three spectroscopic observations, one can always identify a pair of spectra with a shorter temporal separation and another pair with a longer separation, where the two pairs share one common spectrum. When the incident ionizing continuum exhibits stronger variability over the shorter interval, the pattern of variability in the absorption troughs between these spectral pairs becomes closely dependent on the recombination timescale $t_r$, relative to the timescales $\Delta T_\mathrm{short}$ and $\Delta T_\mathrm{long}$.

\begin{itemize}
    \item If $t_r < \Delta T_\mathrm{short} < \Delta T_\mathrm{long}$, the gas has sufficient time to respond to the stronger incident variability in the shorter-interval pair, resulting in more pronounced absorption line variability compared to the longer-interval pair.
    \item In contrast, if $\Delta T_\mathrm{short} < t_r < \Delta T_\mathrm{long}$, the gas cannot fully respond to the changes in the incident light over the short interval. As a result, it becomes possible that the absorption variability is stronger in the longer-interval pair, and the probability of this outcome increases as $t_r$ deviates further from $\Delta T_\mathrm{short}$.
    \item Finally, if $\Delta T_\mathrm{short} < \Delta T_\mathrm{long} < t_r$, the behavior of absorption line variability between the two spectral pairs becomes “chaotic”, depending on the specific pattern of incident ionizing flux and the gas response mechanism. As $t_r$ grows much larger than $\Delta T_\mathrm{long}$, the absorption variability in both pairs tends to converge, and the method becomes ineffective if the maximum observed interval is shorter than $t_r$.
\end{itemize}

Under the condition that the variability in the incident ionizing continuum is stronger over the shorter interval, we define the case in which the absorption line variability is greater in the short-interval pair than in the long-interval pair as a “G1” event. Conversely, if the variability is greater in the long-interval pair, we define it as a “G2” event. As shown above, the probability of observing G1 events decreases monotonically with $t_r$, and can thus be used as an empirical constraint on $t_r$. 
\textbf{The labels “G1” and “G2” are arbitrary and are introduced solely to distinguish these two classes of variability events.}

%当我们获得了单个源的一系列观测光谱时，就相当于在时间轴上对该源设立了一系列观测点，每个观测点上入射到吸收气体的入射光光变是确定的，$\Delta T_\mathrm{short}$ 和 $\Delta T_\mathrm{long}$也是确定的。假如 $t_r$ 会从小到大逐渐变化，那么我们就能想象这些观测点的光变事件就会一个个从G1逐渐向G2变化。不过，实际观测中，吸收线的光变事件和G1概率也是确定的。而从G1事件的概率到 $t_r$具体值的映射关系，我们则通过假设吸收气体对入射光的响应是简单的时间平均，在随机行走模型模拟的AGN光曲线上对这些观测点的行为进行模拟来构建。

When a sequence of spectra is obtained for a given source, it can be used to construct a set of observational points. Each observational point corresponds to a specific combination of short- and long-interval spectral pairs, with a known incident flux variability and measured values of $\Delta T_\mathrm{short}$ and $\Delta T_\mathrm{long}$. 
Assuming that $t_r$ increases gradually, one can imagine that the absorption variability events would transition from G1 to G2 among these observational points. 
In practice, the absorption line variability and the probability of G1 events are determined observationally. The mapping from the observed G1 probability to the corresponding value of $t_r$ is established by simulating the behavior of these observational points on AGN light curves generated using the damped random walk (DRW) model, under the assumption that the gas responds to the incident ionizing flux through simple temporal averaging.

\subsubsection{G1/G2 Event Classification}

%在实验中，我们会对所有三元光谱的所有长短时间间隔光谱对组合进行测试。
%为了保证光谱对之间有明显的光变以及“较短时间间隔的光谱对之间有更强的入射电离连续谱变化”这个前提成立，我们取1500~\AA\处的连续谱流量的相对变化（如下式）来代表入射电离连续谱的光变。然后，我们指选择满足下列条件的三元光谱来进行研究
In the analysis, we examine all possible combinations of short- and long-interval spectral pairs within each spectral triplet.
To ensure (1) significant flux variability between the spectral pairs, and (2) that the short-interval pair exhibits stronger variability in the incident ionizing continuum, we use the relative variation in the continuum flux at 1500~\AA\ as a proxy for ionizing continuum variability:
\begin{equation}
    |\Delta L / L| = \left| \frac{2(L_2 - L_1)}{L_1 + L_2} \right|
\end{equation}
where $L_1$ and $L_2$ denote the continuum fluxes measured at the first and second epochs, respectively.
We then select only those spectral triplets that satisfy the following criteria for further analysis:
\begin{itemize}
    \item $|\Delta L/L|_\mathrm{short} > 0.05$
    \item $|\Delta L/L|_\mathrm{long} > 0.05$
    \item $|\Delta L/L|_\mathrm{short} - |\Delta L/L|_\mathrm{long} > 0.2$
\end{itemize}

For each such triplet, we first evaluated whether significant absorption line variability occurred in the short and long time intervals using the criteria described in Section~\ref{sec:3.1}. 
For each absorption trough in which variability was detected, we calculated the total variability significance, $N_\sigma$, by summing over all variable wavelength segments within the trough. 
Based on this procedure, the criteria for identifying G1 and G2 events were defined as follows:
\begin{itemize}
    \item G1: if $N_\sigma$ of $\Delta T_\mathrm{short}$ exceeds that of $\Delta T_\mathrm{long}$ and $N_\sigma > 3$
    \item G2: if $N_\sigma$ of $\Delta T_\mathrm{long}$ exceeds that of $\Delta T_\mathrm{short}$ and $N_\sigma > 3$
    \item Other cases with no significant variability were excluded from G1/G2 statistics
\end{itemize}
%Finally, for each absorption trough, we computed the fraction of valid events that were classified as G1, yielding $F(\mathrm{G1})$ ($F(\mathrm{G1}) = N_{G1}/(N_{G1}+N_{G2})$), which serves as the key observational constraint on the recombination timescale $t_r$.
Finally, for each absorption trough, we calculated the G1 event fraction, $F(\mathrm{G1}) = N_{G1}/(N_{G1} + N_{G2})$, which serves as the primary observational input for constraining the recombination timescale $t_r$.

The G1/G2 classification outcomes for the three absorption troughs dominated by single velocity components are summarized in Tables~\ref{tab:trough_A} and \ref{tab:trough_I}.

\subsubsection{DRW Simulation of Ionizing Continuum Variability}
\label{sec:drw}

To establish the relationship between the observed G1 event fraction $F(\mathrm{G1})$ and the recombination timescale $t_r$, we simulated the variability of the incident ionizing continuum using a damped random walk (DRW) model, which is a widely adopted statistical framework for describing AGN photometric variability (e.g., \cite{DRW_2009,DRW_2010,MacLeod_2010_DRW,MacLeod_2012_DRW}).

As a well-studied Seyfert galaxy with extensive multi-wavelength, long-term monitoring data, NGC 5548 provides a robust empirical foundation for constructing a realistic DRW model of its ionizing continuum. 
In our analysis, the variability of absorption lines is assumed to be driven by variations in the incident ionizing continuum, with a delayed response governed by the recombination timescale $t_r$. 
We applied the DRW model to simulate the temporal evolution of the ionizing continuum, and incorporated the effect of recombination delay by averaging the continuum flux over a $t_r$-long interval preceding each selected time point.
Previous studies have shown that the variability of BAL equivalent widths increases with time interval and is also consistent with a random-walk model (\cite{Filiz_Ak_2013}).

To calibrate the DRW parameters, we adopted values derived at 1367 Å by \cite{Yu_2020} from the AGN STORM campaign spectra of NGC 5548 in 2014: the damping timescale $\tau = 125$ days and the standard deviation of variability $\sigma = 17.38 \times 10^{-15}$ erg s$^{-1}$ cm$^{-2}$ Å$^{-1}$. 
%Since our analysis is based on relative flux changes $|\Delta L/L|$, the absolute scale of $\sigma$ is irrelevant. 

To approximate the DRW behavior at the C IV ionization threshold (47.9–64.5 eV, or 259–192 Å, corresponding to C III → C IV and C IV → C V transitions), we extrapolated the characteristic timescale to $\sim$200 Å using two empirical relations. The empirical wavelength dependence $\tau \propto \lambda^{2/3}$ (\cite{2017ApJ...847..132G}) yields $\tau \approx 35$ days, while the thermal timescale scaling $\tau \propto \lambda^2$ (\cite{MacLeod_2010_DRW}) gives $\tau \approx 2.7$ days. Taking the geometric mean, we adopted $\tau = 10$ days at 200 Å for our simulations.

%关于 $\mathrm{SF}_\infty$ 的取值，我们先根据$\Delta m = -\frac{2.5}{\ln 10}\frac{\Delta F}{F} \approx -1.086\frac{\Delta F}{F}$将 \cite{Yu_2020} 测得的 $\sigma$转换为星等，然后根据$\mathrm{SF}_\infty = \sqrt{2}\sigma$的关系式得到1367 Å处的$\mathrm{SF}_\infty$。根据\cite{MacLeod_2010_DRW}中的conclusion 和 图 13给出的经验关系式，$\mathrm{SF}_\infty \propto \lambda^{-0.479}$，我们求出了 200 Å处的$\mathrm{SF}_\infty = 1.579 mag$
However, the variability amplitude in the simulations is parameterized by $\mathrm{SF}_\infty$ rather than $\sigma$. To determine the value of $\mathrm{SF}_\infty$, the long-timescale limit of the structure function, we first converted the $\sigma$ value measured by \cite{Yu_2020} into magnitudes using the relation $\Delta m = -\frac{2.5}{\ln 10} \frac{\Delta F}{F} \approx -1.086 \frac{\Delta F}{F}$. Then, applying the relation $\mathrm{SF}_\infty = \sqrt{2}\sigma$, we obtained $\mathrm{SF}_\infty$ at 1367 Å. Based on the empirical relation $\mathrm{SF}_\infty \propto \lambda^{-0.479}$ given in the conclusion and Figure 13 of \cite{MacLeod_2010_DRW}, we extrapolated the value to 200 Å and derived $\mathrm{SF}_\infty = 1.579$ mag. 
Detailed descriptions of the DRW model parameters and their relationships are given in \cite{MacLeod_2010_DRW}.

% 模拟过程
For each spectral triplet associated with variability events, we extracted the short and long time intervals ($\Delta T_\mathrm{short}, \Delta T_\mathrm{long}$) and the corresponding relative flux changes ($|\Delta L/L|_\mathrm{short}, |\Delta L/L|_\mathrm{long}$) from the observational data. These values were then used to identify matching configurations in simulated AGN light curves generated via a DRW model.

To mimic the typical flux uncertainty of AGN STORM spectra, we added a 2\% Gaussian relative flux noise to the entire simulated light curve. A match was accepted if the light curve contained three time points ($t_1$, $t_2$, $t_3$) such that the time intervals matched the observed values, and the relative flux variations satisfied both of the following conditions for each pair:
$|\Delta L/L|_{\mathrm{sim}} - |\Delta L/L|_{\mathrm{obs}}| < 0.01 \quad \text{and} \quad \frac{|\Delta L/L|_{\mathrm{sim}} - |\Delta L/L|_{\mathrm{obs}}|}{|\Delta L/L|_{\mathrm{obs}}} < 0.02$.

Once a matching configuration was found, we simulated different recombination timescales $t_r$ by averaging the continuum flux over a window of length $t_r$ preceding each of the three matched time points, yielding $L_1$, $L_2$, and $L_3$. The resulting $|\Delta L/L|_\mathrm{short} and |\Delta L/L|_\mathrm{long}$ values were used to classify the case as:
\begin{itemize}
    \item G1 if $|\Delta L/L|_\mathrm{short} > |\Delta L/L|_\mathrm{long}$,
    \item G2 if $|\Delta L/L|_\mathrm{short} < |\Delta L/L|_\mathrm{long}$.
\end{itemize}

\begin{figure}
    \centering
    \includegraphics[width=0.75\linewidth]{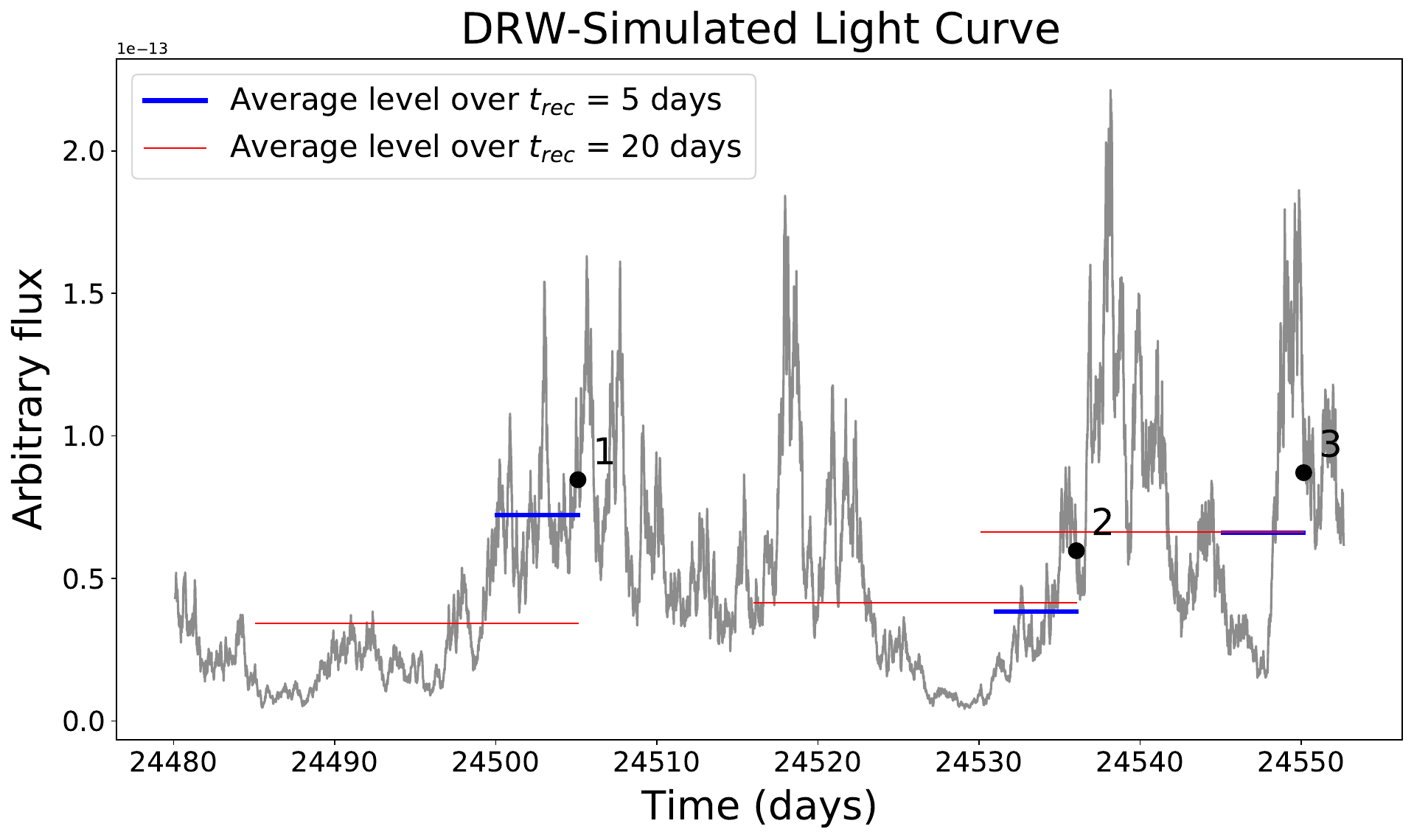}
    \caption{The gray curve shows a continuum light curve generated using the DRW model. The three numbered points mark the epochs matched to the three observed spectra in the simulation, with time intervals of $\Delta T_{12} = 30.9$ days and $\Delta T_{13} = 45.0$ days, corresponding to the short- and long-baseline intervals, respectively. The horizontal lines indicate the mean continuum flux levels within the response windows defined by the recombination timescale $t_{\rm rec}$. For $t_{\rm rec}=5$ days (blue lines), the difference in the mean flux level between epochs 1 and 2 is larger than that between epochs 1 and 3, i.e., $|\Delta L_{12}^{\rm mean}| > |\Delta L_{13}^{\rm mean}|$. Since the continuum fluxes at the three epochs are comparable, it follows that $|\Delta L^{\rm mean}/L|_{12} > |\Delta L^{\rm mean}/L|_{13}$ resulting in a G1 event. 
    In contrast, for a longer recombination timescale of $t_{\rm rec} = 20$ days (red lines), the change in the mean flux level between epochs 1 and 3 becomes larger, $|\Delta L_{13}^{\rm mean}| > |\Delta L_{12}^{\rm mean}|$, leading to $|\Delta L^{\rm mean}/L|_{13} > |\Delta L^{\rm mean}/L|_{12}$ and thus a G2 event.}
    \label{fig:drw_example}
\end{figure}

% 图\ref{fig:drw_example} 给这一过程一个简单示例。
An illustration of this procedure is provided in Figure~\ref{fig:drw_example}.

This procedure was repeated for all observed spectral triplets, and for each assumed $t_r$ (ranging from $\log_{10}(t_r/\mathrm{days})$ = -1.0 to 2.0 in steps of 0.01 dex), we computed the resulting G1 fraction. Running this process 1000 times provided a statistically robust mapping between the G1 occurrence rate and recombination timescale, as shown in Figure~\ref{fig:2}.
\textbf{The lower bound of the sampled $t_r$ range therefore sets the minimum recombination timescale explored in this work. The inferred value for Component 1 lies close to, but remains within, this sampled range.}

% 最后，将从实际光谱中认证好的数据结果带入到模拟出的G1概率和tr的映射关系中，就能从实际吸收坑的光变行为估计出每个速度成分的复合时标tr
%Finally, by applying the validated results from actual spectral data to the simulated mapping between G1 probability and the recombination timescale $t_r$, we can estimate the value of $t_r$ for each velocity component based on the observed variability behavior of the corresponding absorption troughs.
Finally, by applying the identified absorption-line variability events from the observed spectra to the simulated mapping between G1 probability and the recombination timescale $t_r$, we can estimate $t_r$ for each velocity component based on the observed variability behavior of its associated absorption trough.

% 灰色曲线展示了一条通过DRW生成的光变曲线。曲线上的三个数字点标志着模拟时对三条观测光谱的匹配时间点，时间间隔分别为$\Delta T_{12} = 30.9$天和$\Delta T_{13} = 45.0$天，分别对应于短间隔和长间隔。水平线表示由气体复合时标 $t_r$ 定义的响应窗口中连续谱的平均通量水平。对于$t_r = 5$天（蓝线），时期1和2之间的平均通量水平比时期1和3之间的差异更显著，即$|\Delta L_{12}^{\rm mean}| > |\Delta L_{13}^{\rm mean}|$。鉴于三个时期的通量位置相似，可以判断$|\Delta L^{\rm mean}/L|_{12}^{\rm mean} > |\Delta L^{\rm mean}/L|_{13}^{\rm mean}$，预计将发生“G1”事件。相比之下，在更长的复合时标下$t_r = 20$天（红线），第1和第3期之间的平均通量水平变化更大，即$|\Delta L_{13}^{\rm mean}| > |\Delta L_{12}^{\rm mean}|$，导致$|\Delta L^{\rm mean}/L|_{13}^{\rm mean} > |\Delta L^{\rm mean}/L|_{12}^{\rm mean}$，于是此时发生“G2”事件。

\subsection{Outflow Velocity Components and Absorption Trough Identification}
% 我们遵循与前作 \cite{Chen_2026} 相同的外流速度成分划分方案，该方案来自于 \cite{Crenshaw_2003}。
\textbf{We adopt the same outflow velocity component scheme as in our previous study (\cite{Chen_2026}), which is based on the classification of \cite{Crenshaw_2003}.}

\begin{figure}
    \centering
    \includegraphics[width=0.75\linewidth]{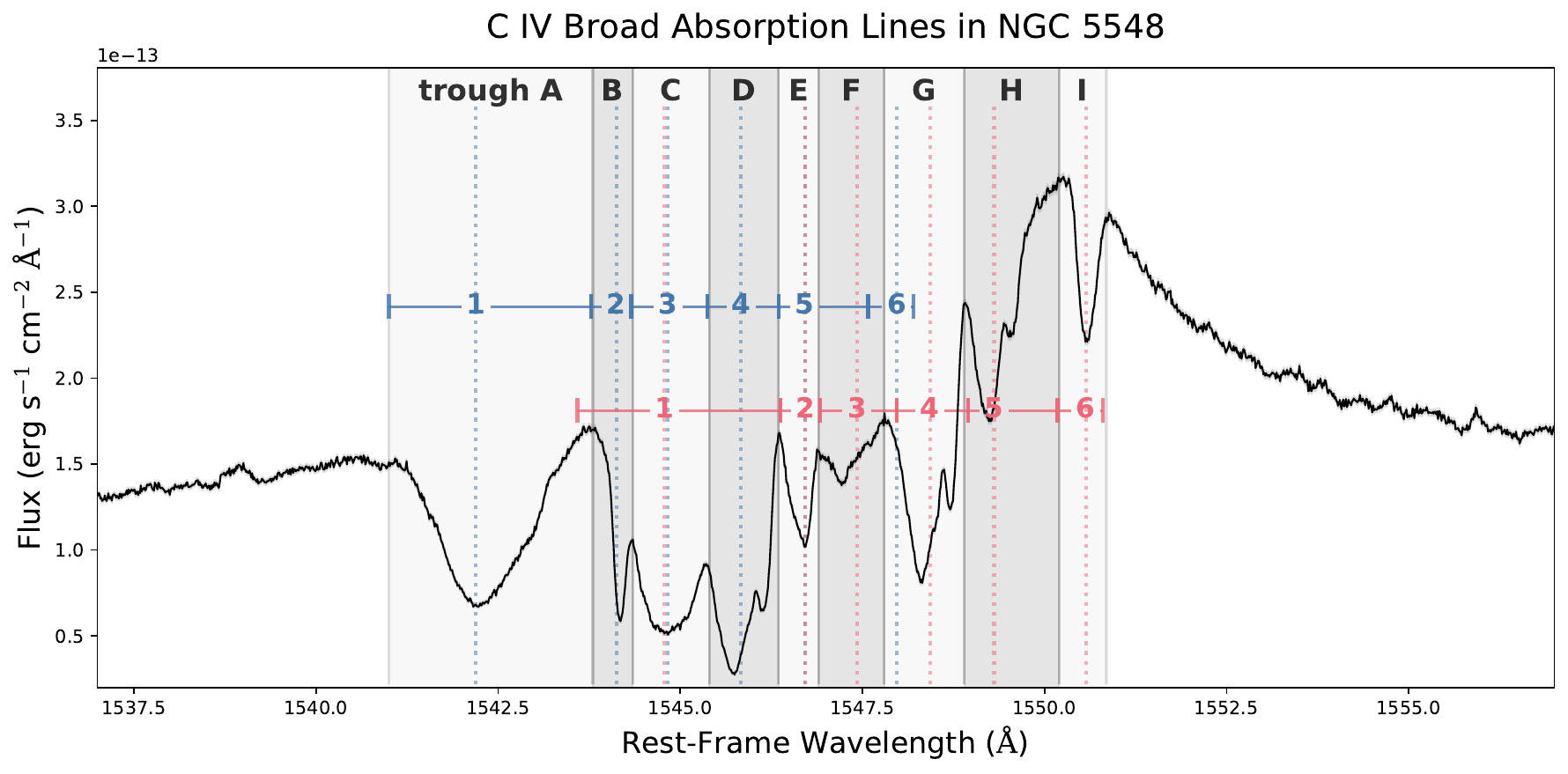}
    \caption{\textbf{Broad C IV absorption lines in NGC 5548.}
    The black curve displays the geometric-mean composite spectrum constructed from all available UV observations. The six ultraviolet outflow components, as defined in \cite{Crenshaw_2003}, are marked by numbered horizontal bars, with red and blue segments denoting the red and blue members of the C IV doublet, respectively. Vertical dashed lines indicate the central wavelengths of each velocity component. Alternating shaded bands highlight the absorption troughs analyzed in this work, with their corresponding labels shown in bold at the top.
    This figure is identical to that used in our previous study (\cite{Chen_2026}), and the same component definitions and segmentation scheme are adopted here for consistency.}
    \label{fig:1}
\end{figure}

% 在光谱中，因为C IV线有红蓝双重线，而红线和蓝线的不同速度成分在波长空间会发生交叠，导致一个吸收坑中有不同外流速度成分的共同影响。因此，我们分析了不同吸收坑中存在的外流速度成分的影响，将其绘制在图 Figure~\ref{fig:1} 中，结果汇总在表 Table~\ref{tab:overlap_Comp}) 中。该图与该表同样来自 \cite{Chen_2026}，为了该文的完整性，在此再次展现。
\textbf{In the C IV absorption profile, the red and blue members of the resonance doublet overlap in wavelength space, causing individual absorption troughs to contain contributions from multiple outflow velocity components. We therefore analyzed the velocity components contributing to each absorption trough. The corresponding component identifications are illustrated in Figure~\ref{fig:1}, with the results summarized in Table~\ref{tab:overlap_Comp}. %Both the figure and the table are reproduced from \citealt{Chen_2026} for completeness.
To make this paper self-contained, both the figure and the table are reproduced from \cite{Chen_2026}.}

% --------------------------------------------------------

\begin{deluxetable}{c|ccc}
\tablecaption{Overlapping Contributions of Multiple Components to Absorption Troughs\label{tab:overlap_Comp}}
\tablehead{
\colhead{Absorption Trough} & \multicolumn{3}{c}{Contributions of Components}
}
\startdata
A & Blue of Comp. 1     & \nodata             & \nodata \\
B & Red of Comp. 1      & Blue of Comp. 2     & \nodata \\
C & Red of Comp. 1      & Blue of Comp. 3     & \nodata \\
D & Red of Comp. 1      & Blue of Comp. 4     & \nodata \\
E & Red of Comp. 2      & Blue of Comp. 5     & \nodata \\
F & Red of Comp. 3      & Blue of Comp. 5     & Blue of Comp. 6 (minor) \\
G & Red of Comp. 4      & Blue of Comp. 6     & Red of Comp. 3 (minor)  \\
H & Red of Comp. 5      & \nodata             & \nodata \\
I & Red of Comp. 6      & \nodata             & \nodata \\
\enddata
\end{deluxetable}

\section{Results}
% 通过对速度成分1、6对应的 trough A 和 I进行吸收线光变认证和对应的G1、G2事件认证，我们测出了对应的这两个成分的 G1 事件概率，分别为 $86.3\pm0.2\%$(18499/21439) 和 $56.9\pm0.5\%$(6490/11416)。
By identifying variability in the absorption troughs A and I corresponding to velocity components 1 and 6, respectively, and associating them with G1 and G2 events, we determine the G1 event probabilities $F_{G1}$ for these two components to be $86.3\pm0.2\%$ (18499/21439) and $56.9\pm0.5\%$ (6490/11416), respectively.

\begin{figure}
    \centering
    \includegraphics[width=0.6\linewidth]{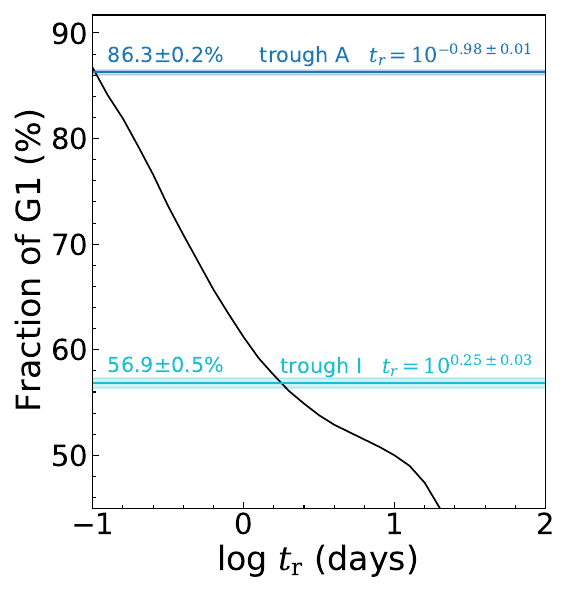}
    \caption{\textbf{Recombination timescale estimates based on the G1 event probability method.} 
    The black curve shows the simulated mapping between $F(\mathrm{G1})$ (the fraction of G1-classified events) and $\log t_r$ (in days), derived from DRW-modeled light curves of the ionizing continuum. Colored horizontal lines represent the measured $F(\mathrm{G1})$ values for different absorption troughs, with shaded bands indicating their 1$\sigma$ uncertainties. The left-hand labels indicate the observed $F(\mathrm{G1})$ values, while the corresponding $t_r$ values inferred from the simulation are shown on the right.}
    \label{fig:2}
\end{figure}

% 随后，将上述概率代入 \ref{sec:drw} 小节叙述的DRW光曲线模拟方法得到的 FG1 和复合时标 tr的映射关系，由此可以估计出成分1和6的复合时标为 $10^{-0.98\pm0.01}$ 和 $10^{0.25\pm0.03}$ day，如图\ref{fig:2}所示。
Substituting these probabilities into the mapping between $F_{G1}$ and the recombination timescale $t_r$ derived from the DRW light curve simulation method described in section~\ref{sec:drw}, we estimate the recombination timescales for components 1 and 6 to be $10^{-0.98\pm0.01}$ and $10^{0.25\pm0.03}$ days, respectively, as shown in Figure~\ref{fig:2}.

% 相比于我们之前的工作中（\cite{Chen_2026}），成分1仅能得出上限（$t_r < 2.509$ days）的结果，这个方法通过数值模拟能约束出近似数值，且与之前的方法的结果相符合。同时，成分6的结果与上一篇文章的相比也相近，说明两个方法互相印证（见\ref{sec:compared}节）。
%\textbf{Compared with our previous study (\cite{Chen_2026}), in which only an upper limit on the recombination timescale of component 1 could be derived ($t_r < 2.509$ days), the numerical simulation approach employed here allows us to estimate its value directly. The resulting timescale is consistent with the previous constraint. Likewise, the recombination timescale derived for component 6 is in good agreement with that reported in the earlier study. This consistency between the two results provides an independent validation of the two methods.}
In our previous study (\cite{Chen_2026}), we estimated the recombination timescales of the velocity components in NGC 5548 using detection probability curves derived from the variability of the C IV absorption troughs. The recombination timescale of component 1 could only be constrained as an upper limit, \textbf{$t_r < 2.51$ days}, whereas that of component 6 was measured to be \textbf{$t_r = 4.83\pm 1.28$ days}. 
In the present work, the numerical simulation approach provides a direct estimate for component 1 that remains consistent with the previous constraint, while the result for component 6 is broadly consistent with the earlier measurement. The agreement between the two sets of results provides an independent validation of the two methods.
%Accordingly, we include the current measurements in the comparison figure of NGC 5548 outflow distance estimates from the literature presented in our earlier work, where they likewise show good agreement (see Figure~\ref{fig:R_comparison}).

% 在该系列的上一篇文章(\cite{Chen_2026})中，我们已经通过 C IV 和 Si IV的吸收坑柱密度和Cloudy（\cite{Cloudy2023}）模拟求出了这两个速度成分的电离参数，这里沿用，分别为 log $U_{\mathrm{H}}$ = $-1.81^{+0.13}_{-0.12}$ 和 $-0.56^{\infty}_{-0.71}$(以及C IV和C V的数密度比 $n_{\mathrm{C\,V}}/n_{\mathrm{C\,IV}} = 0.579^{+0.125}_{-0.046}$和$15.977^{+1.485}_{-0.270}$)。我们选择的SED是D22 (\cite{D22}), which was constructed based on the 2014 observations from the AGN STORM project. 
In our previous study in this series (\cite{Chen_2026}), we derived the ionization parameters of these two components using C IV and Si IV absorption trough column densities and Cloudy simulations (\cite{Cloudy2023}). These values are adopted here: log $U_{\mathrm{H}} = -1.81^{+0.13}_{-0.12}$ for component 1 and $-0.56^{\infty}_{-0.71}$ for component 6, along with the ionic density ratios $n_{\mathrm{C\,V}}/n_{\mathrm{C\,IV}} = 0.579^{+0.125}_{-0.046}$ and $15.977^{+1.485}_{-0.270}$, respectively. The SED we adopt is D22 (\cite{D22}), which was also constructed based on 2014 observations from the AGN STORM project.
%基于 D22 SED和所有光谱几何平均的复合谱的 luminosity at 1500 Å，我们估计出热光度 $L_{\mathrm{bol}} = 2.01 \times 10^{44} \mathrm{erg} \mathrm{s}^{-1}$ 以及电离氢光子发射率 $Q_H = 1.51 \times 10^{54} \mathrm{s}^{-1}$。
Based on the D22 SED and the geometric mean composite spectrum luminosity at 1500 Å, we estimate a bolometric luminosity of $L_{\mathrm{bol}} = 2.01 \times 10^{44}\ \mathrm{erg\ s}^{-1}$ and a hydrogen-ionizing photon emission rate of $Q_H = 1.51 \times 10^{54}\ \mathrm{s}^{-1}$.
% 根据式子 \eqref{eq:1} 和 \eqref{eq:2}，以及误差的不对称性选择蒙特卡洛方法，将上述数值代入后得到两个成分对应的外流半径分别为 $0.77^{+0.10}_{-0.10}$ 和 $1.72^{+1.74}_{-1.72}$ pc。
Using Equations~\eqref{eq:1} and \eqref{eq:2}, we propagate the asymmetric uncertainties of the input parameters through a Monte Carlo approach. The median values of the resulting distributions yield outflow radii of $0.77^{+0.10}_{-0.10}$ and $1.72^{+1.74}_{-1.72}$ pc for components 1 and 6, respectively (see Table~\ref{table:D22+obscured_SED} for results). For component 6, the lower uncertainty formally extends to zero, and we therefore additionally report the corresponding upper-limit interpretation in Table 2.

% 同样的，因为 NGC 5548中可能存在“遮蔽风”（\cite{Kaastra_2014_ngc5548_2013_campaign, Arav_2015}），因此我们还使用 obscured SED （\cite{Mehdipour_2015_ngc5548_2013_campaign} ）进行相同步骤的 Cloudy 模拟和运算。结果见表？。
In addition, considering the potential presence of an “obscuring wind” in NGC 5548 (\cite{Kaastra_2014_ngc5548_2013_campaign, Arav_2015}), we repeated the calculations using the obscured SED from \cite{Mehdipour_2015_ngc5548_2013_campaign}. The results are summarized in Table~\ref{table:D22+obscured_SED}.

\begin{figure}
    \centering
    \includegraphics[width=0.75\linewidth]{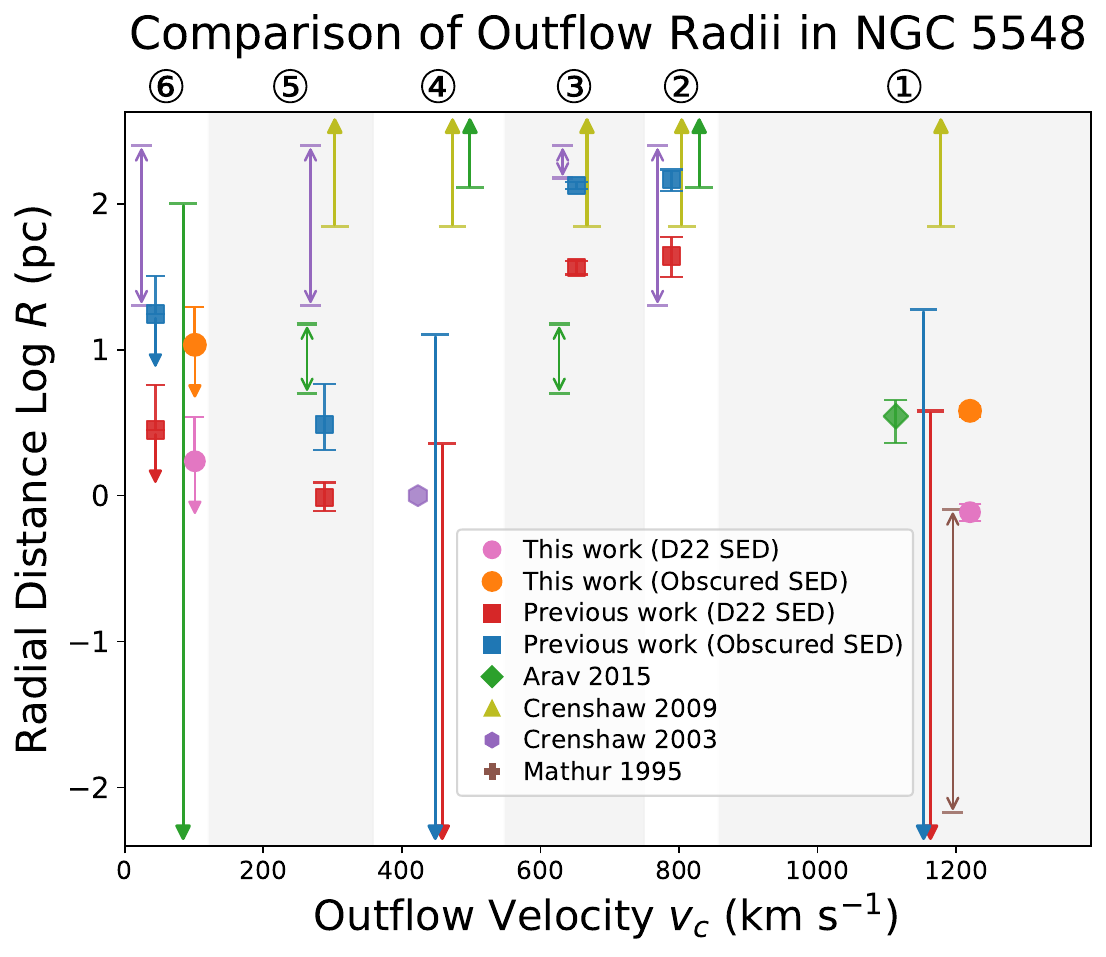}
    \caption{\textbf{Comparison of UV outflow radial distances in NGC 5548 measured in this work, our previous work, and other studies from the literature.} Different point shapes represent different sources: circles for this work, squares for our previous work based on detection probability curves, and other symbols for literature values. Each result is further color-coded by data source. The numbers at the top indicate different velocity components, and the alternating gray and white background bands mark the velocity ranges associated with each outflow component. The x-axis shows velocity ($km\,s^{-1}$), with slight horizontal offsets applied between overlapping results for visual clarity. The y-axis shows the logarithm of the outflow radius in parsecs.}
    \label{fig:R_comparison}
\end{figure}

We also include the current measurements in the comparison figure of NGC 5548 outflow distance estimates from the literature presented in our earlier work, where both the previous and current measurements are shown to be broadly consistent with other distance estimates reported in the literature  (see Figure~\ref{fig:R_comparison}).

%------------------------------------------------------
\begin{deluxetable*}{ccccccccc}
\tablecaption{Physical Parameters of the C\,\textsc{iv} Components in NGC 5548 \label{table:D22+obscured_SED}}
\tablehead{
\colhead{SED} &
\colhead{Component} & \colhead{$v_c$\tablenotemark{a}} & \colhead{Velocity Range} & \colhead{log $U_{\mathrm{H}}$} & \colhead{log $N_{\rm H}$} & \colhead{$\frac{n_{\mathrm{C\,V}}}{n_{\mathrm{C\,IV}}}$\tablenotemark{b}} & \colhead{$Log t_r$\tablenotemark{c}} & \colhead{$R$} \\
\colhead{} & \colhead{} & \colhead{($km\,s^{-1}$)} & \colhead{($km\,s^{-1}$)} & \colhead{} & \colhead{(cm$^{-2}$)} & \colhead{} & \colhead{(days)} & \colhead{(pc)}
}
\startdata
\multirow{2}{*}{D22}
    & 1 & 1162.9 & $1394.2 - 856.0$ & $-1.81^{+0.13}_{-0.12}$ & $19.40^{+0.06}_{-0.06}$ & $0.579^{+0.125}_{-0.046}$ & $10^{-0.98 \pm 0.01}$ & $0.77^{+0.10}_{-0.10}$ \\
    & 6 & 44.3 & $120.0-0.0$ & $-0.56^{+\infty}_{-0.71}$ & $19.44^{+1.34}_{-\infty}$ & $15.977^{+1.485}_{-0.270}$ & $10^{0.25 \pm 0.03}$ & $1.72^{+1.74}_{-1.72}\,(<3.46)$ \\
\hline
\multirow{2}{*}{Obscured SED}
    &1 & 1162.9 & $1394.2-856.0$ & $-2.24^{+0.09}_{-0.08}$ & $19.46^{+0.05}_{-0.06}$ & $0.753^{+0.189}_{-0.052}$ & $10^{-0.98 \pm 0.01}$ & $3.81^{+0.33}_{-0.34}$ \\
    & 6 & 44.3 & $120.0-0.0$ & $-1.25^{+inf}_{-0.60}$ & $19.24^{+1.70}_{-\infty}$ & $14.580^{+0.845}_{-0.205}$ & $10^{0.25 \pm 0.03}$ & $10.80^{+8.79}_{-10.80}\,(<19.59)$ \\
\enddata
\tablecomments{Physical properties of the two UV absorption components in NGC 5548 derived using the D22 and Obscured SEDs.
}
\tablenotetext{a}{Velocity centroid of the component.}
\tablenotetext{b}{The ratio of the number densities of \ion{C}{5} to \ion{C}{4}.}
\tablenotetext{c}{The recombination timescale.}
\end{deluxetable*}
%-------------------------------------------------

\section{Discussion}

\subsection{Impact of Absorption-line Saturation and Other Drivers of Absorption-line Variability}
% 吸收线饱和的影响和其他吸收线光变因素的分析

% 吸收线光变的原因除了本文前提“入射电离连续谱的变化”外，还有覆盖因子的变化/吸收物质整体运动。相关分析已在上一篇文章完成，详见 \cite{Chen_2026} 的 5.1节。结论是这些运动不会是光变时标的主因。
In addition to variations in the incident ionizing continuum, which constitute the underlying assumption of this work, absorption-line variability may also arise from changes in the covering factor or bulk motions of the absorbing material. %These possibilities were investigated in our previous study; a detailed discussion can be found in Section 5.1 of \cite{Chen_2026}.
These possibilities were investigated in our previous study (see Section 5.1 of \cite{Chen_2026}), where we concluded that such motions are unlikely to be the primary driver of the observed variability timescales.

% 关于吸收线饱和的影响，由于吸收坑表现出光变，因此它不会是深度饱和的情况。我们将自己测得的值和前人文献的结果比较，将测得的离子柱密度作为下限来约束电离参数，结论是并不影响我们文章的主要结论。详情可见上篇文章的“5.4. Impact of Column Density and Ionization Parameters on the Results”。
Regarding the impact of absorption-line saturation, the observed variability of the absorption troughs suggests that the lines are unlikely to be heavily saturated. We compared our measured values with those reported in the literature and treated the derived ionic column densities as lower limits when constraining the ionization parameters. 
For component 1, the effect is modest, leading to changes of only a factor of a few in the inferred distance. For component 6, however, the distance R could increase by as much as an order of magnitude. 
A detailed discussion is presented in Section 5.4, “Impact of Column Density and Ionization Parameters on the Results,” of \cite{Chen_2026}.

% 为了估计吸收线饱和的影响，我们同样将测得的离子柱密度设为下限，通过 Cloudy 模拟的结果，求出了电离参数的下限，根据式子2，得到了外流距离的上限。结果汇总在表？中，成分1的影响较小，仅有两三倍的变化幅度；成分6或许会增大一个数量级。
%To estimate the impact of absorption line saturation, we also treat the measured ionic column densities as lower limits. Using Cloudy simulations, we then obtain lower limits on the ionization parameters, and consequently, upper limits on the outflow distances based on equations~\eqref{eq:2}. The results are summarized in Table[?]. 
%For component 1, the effect is relatively small, with changes within a factor of a few, while for component 6, the distance could increase by up to an order of magnitude.
%As discussed in our previous study (\cite{Chen_2026}), absorption line saturation may lead to underestimates of ionic column densities. By treating these values as lower limits, we estimated corresponding upper limits on $R$ using Cloudy simulations. For component 1, the effect is minor (a factor of a few), while for component 6, $R$ may be underestimated by up to an order of magnitude.

% Comparison of UV outflow radial distances in NGC 5548 measured in this work and previous work and 文献中的结果。图中不同点形代表不同工作（圆形为this work；方形是previos work，用探测率曲线测量的结果），且用不同颜色将每种结果进行了区分。图片上方的编号代表的不同速度成分，背景灰白交错的色块代表不同外流成分的范围。横坐标是速度（km/s），相同成分间速度有偏移使其互相错开；纵坐标是外流距离 Log R in pc。

% -------------------------------------------------------------------

% -------------------------------------------------------------------
\subsection{Impact of DRW Timescale Assumptions}
\label{sec:5.1}
\begin{figure}
    \centering
    \includegraphics[width=0.6\linewidth]{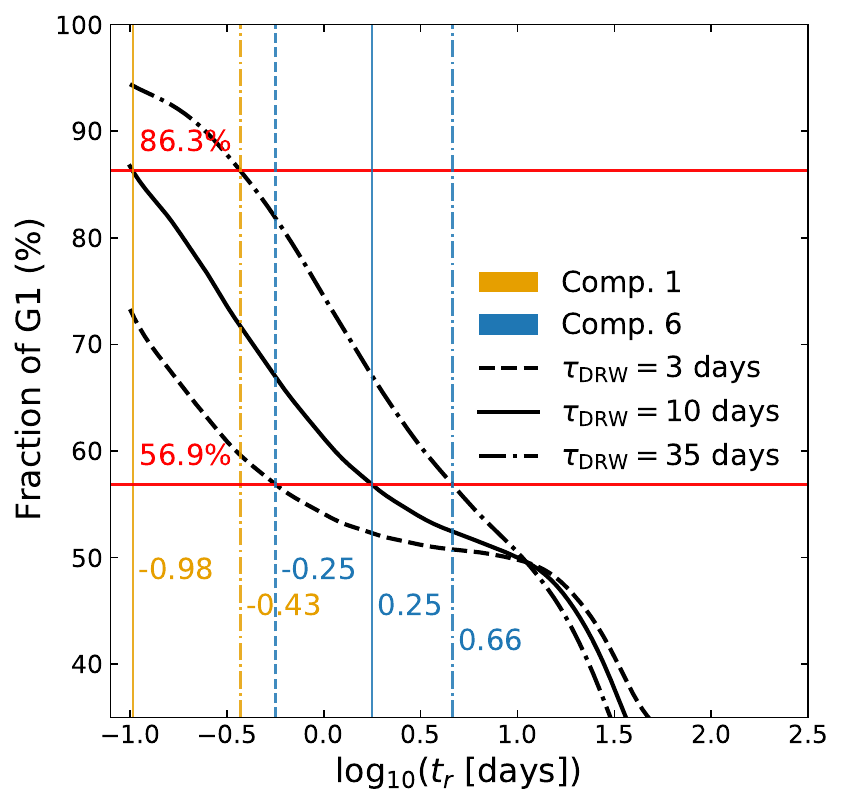}
    \caption{\textbf{Systematic test of the effect of varying the characteristic timescale parameter in the DRW simulations.} The three curves show the mapping between $F(\mathrm{G1})$ (the fraction of G1-classified events) and $\log t_r$ (in days), based on simulated light curves generated with DRW characteristic timescales of 3, 10, and 35 days. The fiducial value adopted in this work is $\tau_{\rm DRW} = 10$ days. The red horizontal lines indicate the observed G1 event fractions measured from the absorption troughs corresponding to the two velocity components. Vertical lines mark the inferred $\log t_r$ values from the intersection with each curve: \textbf{the yellow lines correspond to component 1, and the blue lines to component 6.}}
    \label{fig:drw_3_10_35day}
\end{figure}

% 5.1
% DRW模拟中（\ref{sec:drw} 小节）我们取了10day的特征时标来生成光曲线，为了展示它变化带来的结果变动，我们还取了经验关系中估计出来的其它数值（$\tau = 3$ 和 35 day）进行了模拟，结果如图？所示。
In the DRW simulations described in section~\ref{sec:drw}, we adopted a characteristic damping timescale of 10 days to generate the simulated light curves. To assess how this choice affects the resulting recombination timescales, we also performed simulations using two alternative values estimated from empirical relations: $\tau_{\rm DRW} = 3$ and 35 days. The results are shown in Figure~\ref{fig:drw_3_10_35day}.

% 成分1 用正文$\tau_{\rm DRW} = 10$ day的配置测出的复合时标为 $10^{-0.98\pm0.01}$ day；在 $\tau_{\rm DRW} = 35$ day的模拟下，复合时标 $t_r = 10^{-0.43\pm0.01}$ ($0.37 \pm 0.01$) day；在 $\tau_{\rm DRW} = 3$ day的模拟下，超出了模拟的精度范围，但从趋势上看，$t_r < 0.1$ day。
For component 1, the recombination timescale derived under the baseline configuration ($\tau_{\rm DRW} = 10$ days) is $10^{-0.98\pm0.01}$ days. 
With $\tau_{\rm DRW} = 35$ days, the recombination timescale becomes $t_r = 10^{-0.43\pm0.01}$ days. 
When using $\tau_{\rm DRW} = 3$ days, $t_r$ falls below the resolution limit of our simulations, but the trend suggests $t_r < 0.1$ days (see the dotted curve corresponding to $\tau_{\rm DRW} = 3$ days in Figure~\ref{fig:drw_3_10_35day}).

% 同理，成分6 用正文$\tau_{\rm DRW} = 10$ day的配置测出的复合时标为 $10^{0.25\pm0.03}$ ($1.78 \pm 0.13$) day；在 $\tau_{\rm DRW} = 35$ day的模拟下，复合时标 $t_r = 10^{0.66\pm0.02}$ ($4.62 \pm 0.23$) day；在 $\tau_{\rm DRW} = 3$ day的模拟下，复合时标 $t_r = 10^{-0.25\pm0.04}$ ($0.57 \pm 0.06$) day。
Similarly, for component 6, the baseline configuration yields a recombination timescale of $10^{0.25\pm0.03}$ days. 
With $\tau_{\rm DRW} = 35$ days, the result is $t_r = 10^{0.66\pm0.02}$  days, 
while for $\tau_{\rm DRW} = 3$ days, $t_r = 10^{-0.25\pm0.04}$  days.

% 可以看出，在两种参数的取法上，都与$\tau_{\rm DRW} = 10$ day的结果不到1个数量级的差距。如果再结合之前探测率曲线测出的结果进行比较（见下节），特征时标偏向35 day的一方更符合。
In both cases, the differences from the $\tau_{\rm DRW} = 10$ day results are within one order of magnitude. 
% 从 \label{eq:1} 和 \label{eq:2} 中可以推出，$\Delta \log R = 2\Delta \log t_r$，因此 $\tau_{\rm DRW}$ 带来的误差并不改变我们之前的结论。
\textbf{Since Equations~\eqref{eq:1} and \eqref{eq:2} imply $\Delta\log R = 2\,\Delta\log t_r$, the uncertainty introduced by the choice of $\tau_{\rm DRW}$ does not alter our main conclusions.}
When compared with the detection probability curve–based analysis (\cite{Chen_2026}) , the simulation results appear more consistent under the assumption of a larger characteristic timescale (e.g., $\tau_{\rm DRW} = 35$ days).

% 变换DRW模拟中生成光曲线的特征时标参数观察影响。图中三条曲线就是在3、10、35day的特征时标下模拟出来的 $F(\mathrm{G1})$ (the fraction of G1-classified events) and $\log t_r$ (in days)的映射关系曲线。红色横线代表两个速度成分对应的吸收坑下测出的G1事件比例；竖线标出了它们在三条模拟曲线（$\tau =$ 3、10、35day）上映射出来的 $\log t_r$，其中蓝色竖线是成分1的结果，紫色竖线是成分6的对应结果。

\section{Conclusion}
In this study, we measured the radial distances of UV absorption outflow components in NGC 5548 using the G1 event probability method. 
This analysis is based on multi-epoch HST/COS spectroscopic observations from the 2014 AGN STORM campaign, supplemented by earlier data from 2013. 
%We focused specifically on absorption troughs dominated by a single velocity component, which allowed us to constrain the recombination timescales ($t_r$) and radial distances ($R$) for components 1 and 6. 
We focused specifically on absorption troughs that are dominated by a single velocity component and exhibit a statistically significant correlation between continuum luminosity and equivalent width, which allowed us to constrain the recombination timescales ($t_r$) and radial distances ($R$) for components 1 and 6.
The $t_r$ values were derived by computing the occurrence probabilities of G1-type absorption-line variability events and applying the mapping between $F(\mathrm{G1})$ and $t_r$ derived from Damped Random Walk simulations of the ionizing continuum.
These recombination timescales, combined with photoionization modeling using CLOUDY, were used to determine the radial distances of the outflow components.
%最后，通过这个方法，我们测出了成分1和6的外流距离为 $0.77^{+0.10}_{-0.10}$ and $1.72^{+1.74}_{-1.72}$ pc，并且结果与前一篇使用探测率曲线测出的结果相近
Finally, using this method, we determined the outflow distances for components 1 and 6 to be $0.77^{+0.10}_{-0.10}$ and $1.72^{+1.74}_{-1.72}$ pc, respectively, which are consistent with the results obtained in our previous work using detection probability curves.
% 生成 DRW 光曲线的参数 $\tau_{\rm DRW}$ 的数值有影响，但并不改变我们的结论。
\textbf{The adopted value of $\tau_{\rm DRW}$ affects the inferred outflow distances to some extent,  but the resulting variations do not alter the main conclusions of this work.}

\begin{table}[htbp]
\centering
\caption[]{\textbf{Sample G1/G2 Event Classifications in Trough A (partial)}. The subscripts "short" and "long" represent the short and long observation intervals, respectively. The full table is provided in Extended Data. \label{tab:trough_A}}
\begin{tabular}{cccccccc}
\hline
No. & $\Delta T_{\rm short}$ (days) & $|\Delta L/L|_{\rm short}$ & $N_{\sigma,\rm short}$ & $\Delta T_{\rm long}$ (days) & $|\Delta L/L|_{\rm long}$ & $N_{\sigma,\rm long}$ & Type of Trough A \\
\hline
1 & 213.8 & 0.576 & 28.05 & 221.6 & 0.302 & 10.62 & G1 \\
2 & 43.4 & 0.287 & 30.08 & 260.3 & 0.081 & 15.36 & G1 \\
3 & 18.0 & 0.585 & 31.56 & 47.2 & 0.224 & 14.10 & G1 \\
4 & 12.1 & 0.610 & 27.15 & 51.1 & 0.315 & 24.05 & G1 \\
5 & 5.4 & 0.402 & 17.85 & 24.3 & 0.190 & 4.50 & G1 \\
\dots & \dots & \dots & \dots & \dots & \dots & \dots & \dots \\
21439 & 3.6 & 0.293 & 18.50 & 6.9 & 0.064 & 11.75 & G1 \\
\hline
\multicolumn{2}{l}{Trough A} Fraction of G1 (\%) :& & & & & & 86.3 $\pm$ 0.2\\
\hline
\end{tabular}
\end{table}

\begin{table}[htbp]
\centering
\caption[]{\textbf{Sample G1/G2 Event Classifications in Trough I (partial)}. The subscripts "short" and "long" represent the short and long observation intervals, respectively. The full table is provided in Extended Data.\label{tab:trough_I}}
\begin{tabular}{cccccccc}
\hline
No. & $\Delta T_{\rm short}$ (days) & $|\Delta L/L|_{\rm short}$ & $N_{\sigma,\rm short}$ & $\Delta T_{\rm long}$ (days) & $|\Delta L/L|_{\rm long}$ & $N_{\sigma,\rm long}$ & Type of Trough I \\
\hline
1 & 11.3 & 0.487 & 3.74 & 49.9 & 0.257 & 0.00 & G1 \\
2 & 19.0 & 0.692 & 0.00 & 55.7 & 0.381 & 6.74 & G2 \\
3 & 9.6 & 0.354 & 6.24 & 39.5 & 0.073 & 4.53 & G1 \\
4 & 36.5 & 0.384 & 3.56 & 59.9 & 0.175 & 5.02 & G2 \\
5 & 14.1 & 0.645 & 7.00 & 45.3 & 0.216 & 0.00 & G1 \\
\dots & \dots & \dots & \dots & \dots & \dots & \dots & \dots \\
11416 & 3.6 & 0.293 & 0.00 & 6.9 & 0.064 & 4.58 & G2 \\
\hline
\multicolumn{2}{l}{Trough I} Fraction of G1 (\%) :& & & & & & 56.9 $\pm$ 0.5\\
\hline
\end{tabular}
\end{table}

%\appendix
\section{acknowledgments}
This work was supported by the National SKA Program of China (No. 2025SKA0130100).
We acknowledge the research grants from the Ministry of Science and Technology of China (National Key Program for Science and Technology Research and Development, No. 2023YFA1608100), the research grants from the China Manned Space Project (CMS-CSST-2025-A08), the National Natural Science Foundation of China (Nos. 12273036, 12222304, 12192220 and 12192221).
This research is based on observations made with the NASA/ESA Hubble Space Telescope obtained from the Space Telescope Science Institute, which is operated by the Association of Universities for Research in Astronomy, Inc., under NASA contract NAS 5–26555. These observations are associated with programs GO-13330 and GO-13184. The HST data presented in this article were obtained from the Mikulski Archive for Space Telescopes (MAST) at the Space Telescope Science Institute. The specific observations analyzed can be accessed via \dataset[doi: 10.17909/mmy4-vz85]{https://doi.org/10.17909/mmy4-vz85}.

\bibliography{refs}{}

@ARTICLE{1969Natur.223..690L,
       author = {{Lynden-Bell}, D.},
        title = "{Galactic Nuclei as Collapsed Old Quasars}",
      journal = {\nat},
         year = 1969,
        month = aug,
       volume = {223},
       number = {5207},
        pages = {690-694},
          doi = {10.1038/223690a0},
       adsurl = {https://ui.adsabs.harvard.edu/abs/1969Natur.223..690L}
}

@ARTICLE{1984ARA&A..22..471R,
       author = {{Rees}, Martin J.},
        title = "{Black Hole Models for Active Galactic Nuclei}",
      journal = {\araa},
         year = 1984,
        month = jan,
       volume = {22},
        pages = {471-506},
          doi = {10.1146/annurev.aa.22.090184.002351},
       adsurl = {https://ui.adsabs.harvard.edu/abs/1984ARA&A..22..471R}
}

@ARTICLE{2005MNRAS.361..776S,
       author = {{Springel}, Volker and {Di Matteo}, Tiziana and {Hernquist}, Lars},
        title = "{Modelling feedback from stars and black holes in galaxy mergers}",
      journal = {\mnras},
         year = 2005,
        month = aug,
       volume = {361},
       number = {3},
        pages = {776-794},
          doi = {10.1111/j.1365-2966.2005.09238.x},
archivePrefix = {arXiv},
       eprint = {astro-ph/0411108},
 primaryClass = {astro-ph},
       adsurl = {https://ui.adsabs.harvard.edu/abs/2005MNRAS.361..776S}
}

@ARTICLE{2006MNRAS.370..645B,
       author = {{Bower}, R.~G. and {Benson}, A.~J. and {Malbon}, R. and {Helly}, J.~C. and {Frenk}, C.~S. and {Baugh}, C.~M. and {Cole}, S. and {Lacey}, C.~G.},
        title = "{Breaking the hierarchy of galaxy formation}",
      journal = {\mnras},
         year = 2006,
        month = aug,
       volume = {370},
       number = {2},
        pages = {645-655},
          doi = {10.1111/j.1365-2966.2006.10519.x},
archivePrefix = {arXiv},
       eprint = {astro-ph/0511338},
 primaryClass = {astro-ph},
       adsurl = {https://ui.adsabs.harvard.edu/abs/2006MNRAS.370..645B}
}

@ARTICLE{2006ApJS..163....1H,
       author = {{Hopkins}, Philip F. and {Hernquist}, Lars and {Cox}, Thomas J. and {Di Matteo}, Tiziana and {Robertson}, Brant and {Springel}, Volker},
        title = "{A Unified, Merger-driven Model of the Origin of Starbursts, Quasars, the Cosmic X-Ray Background, Supermassive Black Holes, and Galaxy Spheroids}",
      journal = {\apjs},
         year = 2006,
        month = mar,
       volume = {163},
       number = {1},
        pages = {1-49},
          doi = {10.1086/499298},
archivePrefix = {arXiv},
       eprint = {astro-ph/0506398},
 primaryClass = {astro-ph},
       adsurl = {https://ui.adsabs.harvard.edu/abs/2006ApJS..163....1H}
}

@ARTICLE{2008MNRAS.390.1399B,
       author = {{Bower}, R.~G. and {McCarthy}, I.~G. and {Benson}, A.~J.},
        title = "{The flip side of galaxy formation: a combined model of galaxy formation and cluster heating}",
      journal = {\mnras},
         year = 2008,
        month = nov,
       volume = {390},
       number = {4},
        pages = {1399-1410},
          doi = {10.1111/j.1365-2966.2008.13869.x},
archivePrefix = {arXiv},
       eprint = {0808.2994},
 primaryClass = {astro-ph},
       adsurl = {https://ui.adsabs.harvard.edu/abs/2008MNRAS.390.1399B}
}

@ARTICLE{2007ApJ...665..120A,
       author = {{Aller}, M.~C. and {Richstone}, D.~O.},
        title = "{Host Galaxy Bulge Predictors of Supermassive Black Hole Mass}",
      journal = {\apj},
         year = 2007,
        month = aug,
       volume = {665},
       number = {1},
        pages = {120-156},
          doi = {10.1086/519298},
       adsurl = {https://ui.adsabs.harvard.edu/abs/2007ApJ...665..120A}
}

@ARTICLE{2008MNRAS.391..481S,
       author = {{Somerville}, Rachel S. and {Hopkins}, Philip F. and {Cox}, Thomas J. and {Robertson}, Brant E. and {Hernquist}, Lars},
        title = "{A semi-analytic model for the co-evolution of galaxies, black holes and active galactic nuclei}",
      journal = {\mnras},
         year = 2008,
        month = dec,
       volume = {391},
       number = {2},
        pages = {481-506},
          doi = {10.1111/j.1365-2966.2008.13805.x},
archivePrefix = {arXiv},
       eprint = {0808.1227},
 primaryClass = {astro-ph},
       adsurl = {https://ui.adsabs.harvard.edu/abs/2008MNRAS.391..481S}
}

@ARTICLE{2010MNRAS.406..822M,
       author = {{McCarthy}, I.~G. and {Schaye}, J. and {Ponman}, T.~J. and {Bower}, R.~G. and {Booth}, C.~M. and {Dalla Vecchia}, C. and {Crain}, R.~A. and {Springel}, V. and {Theuns}, T. and {Wiersma}, R.~P.~C.},
        title = "{The case for AGN feedback in galaxy groups}",
      journal = {\mnras},
         year = 2010,
        month = aug,
       volume = {406},
       number = {2},
        pages = {822-839},
          doi = {10.1111/j.1365-2966.2010.16750.x},
archivePrefix = {arXiv},
       eprint = {0911.2641},
 primaryClass = {astro-ph.CO},
       adsurl = {https://ui.adsabs.harvard.edu/abs/2010MNRAS.406..822M}
}

@ARTICLE{2012ApJ...751..107B,
       author = {{Borguet}, Benoit C.~J. and {Edmonds}, Doug and {Arav}, Nahum and {Dunn}, Jay and {Kriss}, Gerard A.},
        title = "{A 10 kpc Scale Seyfert Galaxy Outflow: HST/COS Observations of IRAS F22456-5125}",
      journal = {\apj},
         year = 2012,
        month = jun,
       volume = {751},
       number = {2},
          eid = {107},
        pages = {107},
          doi = {10.1088/0004-637X/751/2/107},
archivePrefix = {arXiv},
       eprint = {1205.0189},
 primaryClass = {astro-ph.CO},
       adsurl = {https://ui.adsabs.harvard.edu/abs/2012ApJ...751..107B}
}

@ARTICLE{2014MNRAS.444.1893C,
       author = {{Capellupo}, D.~M. and {Hamann}, F. and {Barlow}, T.~A.},
        title = "{A variable P v broad absorption line and quasar outflow energetics}",
      journal = {\mnras},
         year = 2014,
        month = oct,
       volume = {444},
       number = {2},
        pages = {1893-1900},
          doi = {10.1093/mnras/stu1502},
archivePrefix = {arXiv},
       eprint = {1407.7532},
 primaryClass = {astro-ph.GA},
       adsurl = {https://ui.adsabs.harvard.edu/abs/2014MNRAS.444.1893C}
}

@ARTICLE{2014ApJ...783...58L,
       author = {{Lucy}, Adrian B. and {Leighly}, Karen M. and {Terndrup}, Donald M. and {Dietrich}, Matthias and {Gallagher}, Sarah C.},
        title = "{Tracing the Outflow of a z = 0.334 FeLoBAL: New Constraints from Low-ionization Absorbers in FBQS J1151+3822}",
      journal = {\apj},
         year = 2014,
        month = mar,
       volume = {783},
       number = {1},
          eid = {58},
        pages = {58},
          doi = {10.1088/0004-637X/783/1/58},
archivePrefix = {arXiv},
       eprint = {1401.0605},
 primaryClass = {astro-ph.GA},
       adsurl = {https://ui.adsabs.harvard.edu/abs/2014ApJ...783...58L}
}

@ARTICLE{2011ApJ...728...94L,
       author = {{Leighly}, Karen M. and {Dietrich}, Matthias and {Barber}, Sara},
        title = "{The Discovery of the First He I{\ensuremath{\lambda}}10830 Broad Absorption Line Quasar}",
      journal = {\apj},
         year = 2011,
        month = feb,
       volume = {728},
       number = {2},
          eid = {94},
        pages = {94},
          doi = {10.1088/0004-637X/728/2/94},
       adsurl = {https://ui.adsabs.harvard.edu/abs/2011ApJ...728...94L}
}

@ARTICLE{Arav_Liu_2018,
       author = {{Arav}, Nahum and {Liu}, Guilin and {Xu}, Xinfeng and {Stidham}, James and {Benn}, Chris and {Chamberlain}, Carter},
        title = "{Evidence that 50\% of BALQSO Outflows Are Situated at Least 100 pc from the Central Source}",
      journal = {\apj},
         year = 2018,
        month = apr,
       volume = {857},
       number = {1},
          eid = {60},
        pages = {60},
          doi = {10.3847/1538-4357/aab494},
archivePrefix = {arXiv},
       eprint = {1805.01543},
 primaryClass = {astro-ph.GA},
       adsurl = {https://ui.adsabs.harvard.edu/abs/2018ApJ...857...60A}
}

@ARTICLE{2018ApJ...866....7L,
       author = {{Leighly}, Karen M. and {Terndrup}, Donald M. and {Gallagher}, Sarah C. and {Richards}, Gordon T. and {Dietrich}, Matthias},
        title = "{The z = 0.54 LoBAL Quasar SDSS J085053.12+445122.5. I. Spectral Synthesis Analysis Reveals a Massive Outflow}",
      journal = {\apj},
         year = 2018,
        month = oct,
       volume = {866},
       number = {1},
          eid = {7},
        pages = {7},
          doi = {10.3847/1538-4357/aadee6},
archivePrefix = {arXiv},
       eprint = {1808.02441},
 primaryClass = {astro-ph.GA},
       adsurl = {https://ui.adsabs.harvard.edu/abs/2018ApJ...866....7L}
}

@ARTICLE{2019MNRAS.483.1808H,
       author = {{Hamann}, Fred and {Herbst}, Hanna and {Paris}, Isabelle and {Capellupo}, Daniel},
        title = "{On the structure and energetics of quasar broad absorption-line outflows}",
      journal = {\mnras},
         year = 2019,
        month = feb,
       volume = {483},
       number = {2},
        pages = {1808-1828},
          doi = {10.1093/mnras/sty2900},
archivePrefix = {arXiv},
       eprint = {1810.03686},
 primaryClass = {astro-ph.GA},
       adsurl = {https://ui.adsabs.harvard.edu/abs/2019MNRAS.483.1808H}
}

@ARTICLE{Arav_2012,
       author = {{Arav}, N. and {Edmonds}, D. and {Borguet}, B. and {Kriss}, G.~A. and {Kaastra}, J.~S. and {Behar}, E. and {Bianchi}, S. and {Cappi}, M. and {Costantini}, E. and {Detmers}, R.~G. and {Ebrero}, J. and {Mehdipour}, M. and {Paltani}, S. and {Petrucci}, P.~O. and {Pinto}, C. and {Ponti}, G. and {Steenbrugge}, K.~C. and {de Vries}, C.~P.},
        title = "{Multiwavelength campaign on Mrk 509. X. Lower limit on the distance of the absorber from HST COS and STIS spectroscopy}",
      journal = {\aap},
         year = 2012,
        month = aug,
       volume = {544},
          eid = {A33},
        pages = {A33},
          doi = {10.1051/0004-6361/201118501},
archivePrefix = {arXiv},
       eprint = {1205.2559},
 primaryClass = {astro-ph.CO},
       adsurl = {https://ui.adsabs.harvard.edu/abs/2012A&A...544A..33A}
}

@ARTICLE{Arav_2015,
       author = {{Arav}, N. and {Chamberlain}, C. and {Kriss}, G.~A. and {Kaastra}, J.~S. and {Cappi}, M. and {Mehdipour}, M. and {Petrucci}, P. -O. and {Steenbrugge}, K.~C. and {Behar}, E. and {Bianchi}, S. and {Boissay}, R. and {Branduardi-Raymont}, G. and {Costantini}, E. and {Ely}, J.~C. and {Ebrero}, J. and {di Gesu}, L. and {Harrison}, F.~A. and {Kaspi}, S. and {Malzac}, J. and {De Marco}, B. and {Matt}, G. and {Nandra}, K.~P. and {Paltani}, S. and {Peterson}, B.~M. and {Pinto}, C. and {Ponti}, G. and {Pozo Nu{\~n}ez}, F. and {De Rosa}, A. and {Seta}, H. and {Ursini}, F. and {de Vries}, C.~P. and {Walton}, D.~J. and {Whewell}, M.},
        title = "{Anatomy of the AGN in NGC 5548. II. The spatial, temporal, and physical nature of the outflow from HST/COS Observations}",
      journal = {\aap},
         year = 2015,
        month = may,
       volume = {577},
          eid = {A37},
        pages = {A37},
          doi = {10.1051/0004-6361/201425302},
archivePrefix = {arXiv},
       eprint = {1411.2157},
 primaryClass = {astro-ph.GA},
       adsurl = {https://ui.adsabs.harvard.edu/abs/2015A&A...577A..37A}
}

@ARTICLE{Ebrero_2016,
       author = {{Ebrero}, J. and {Kaastra}, J.~S. and {Kriss}, G.~A. and {Di Gesu}, L. and {Costantini}, E. and {Mehdipour}, M. and {Bianchi}, S. and {Cappi}, M. and {Boissay}, R. and {Branduardi-Raymont}, G. and {Petrucci}, P. -O. and {Ponti}, G. and {Pozo N{\'u}{\~n}ez}, F. and {Seta}, H. and {Steenbrugge}, K.~C. and {Whewell}, M.},
        title = "{Anatomy of the AGN in NGC 5548. VI. Long-term variability of the warm absorber}",
      journal = {\aap},
         year = 2016,
        month = mar,
       volume = {587},
          eid = {A129},
        pages = {A129},
          doi = {10.1051/0004-6361/201527808},
archivePrefix = {arXiv},
       eprint = {1601.02385},
 primaryClass = {astro-ph.HE},
       adsurl = {https://ui.adsabs.harvard.edu/abs/2016A&A...587A.129E}
}

@ARTICLE{He_2022,
       author = {{He}, Zhicheng and {Liu}, Guilin and {Wang}, Tinggui and {Mou}, Guobin and {Green}, Richard and {Bian}, Weihao and {Wang}, Huiyuan and {Ho}, Luis C. and {Sun}, Mouyuan and {Shen}, Lu and {Arav}, Nahum and {Chen}, Chen and {Wu}, Qingwen and {Guo}, Hengxiao and {Lin}, Zesen and {Li}, Junyao and {Yi}, Weimin},
        title = "{Evidence for quasar fast outflows being accelerated at the scale of tens of parsecs}",
      journal = {Science Advances},
         year = 2022,
        month = feb,
       volume = {8},
       number = {6},
          eid = {eabk3291},
        pages = {eabk3291},
          doi = {10.1126/sciadv.abk3291},
archivePrefix = {arXiv},
       eprint = {2202.06227},
 primaryClass = {astro-ph.GA},
       adsurl = {https://ui.adsabs.harvard.edu/abs/2022SciA....8.3291H}
}

@ARTICLE{He_2019,
       author = {{He}, Zhicheng and {Wang}, Tinggui and {Liu}, Guilin and {Wang}, Huiyuan and {Bian}, Weihao and {Tchernyshyov}, Kirill and {Mou}, Guobin and {Xu}, Youhua and {Zhou}, Hongyan and {Green}, Richard and {Xu}, Jun},
        title = "{The properties of broad absorption line outflows based on a large sample of quasars}",
      journal = {Nature Astronomy},
         year = 2019,
        month = jan,
       volume = {3},
        pages = {265},
          doi = {10.1038/s41550-018-0669-8},
archivePrefix = {arXiv},
       eprint = {1812.08982},
 primaryClass = {astro-ph.GA},
       adsurl = {https://ui.adsabs.harvard.edu/abs/2019NatAs...3..265H}
}

@ARTICLE{2017A&A...607A.100M,
       author = {{Mao}, Junjie and {Kaastra}, J.~S. and {Mehdipour}, M. and {Raassen}, A.~J.~J. and {Gu}, Liyi and {Miller}, J.~M.},
        title = "{Density diagnostics of ionized outflows in active galactic nuclei. X-ray and UV absorption lines from metastable levels in Be-like to C-like ions}",
      journal = {\aap},
         year = 2017,
        month = nov,
       volume = {607},
          eid = {A100},
        pages = {A100},
          doi = {10.1051/0004-6361/201731378},
archivePrefix = {arXiv},
       eprint = {1707.09552},
 primaryClass = {astro-ph.HE},
       adsurl = {https://ui.adsabs.harvard.edu/abs/2017A&A...607A.100M}
}

@ARTICLE{2013A&A...551L...6K,
       author = {{Kollatschny}, W. and {Zetzl}, M.},
        title = "{Accretion disk wind as explanation for the broad-line region structure in NGC 5548}",
      journal = {\aap},
         year = 2013,
        month = mar,
       volume = {551},
          eid = {L6},
        pages = {L6},
          doi = {10.1051/0004-6361/201220923},
archivePrefix = {arXiv},
       eprint = {1301.7704},
 primaryClass = {astro-ph.CO},
       adsurl = {https://ui.adsabs.harvard.edu/abs/2013A&A...551L...6K}
}

@ARTICLE{Mehdipour_2024,
       author = {{Mehdipour}, Missagh and {Kriss}, Gerard A. and {Kaastra}, Jelle S. and {Costantini}, Elisa and {Gu}, Liyi and {Landt}, Hermine and {Mao}, Junjie and {Rogantini}, Daniele},
        title = "{First High-resolution Spectroscopy of X-Ray Absorption Lines in the Obscured State of NGC 5548}",
      journal = {\apj},
         year = 2024,
        month = feb,
       volume = {962},
       number = {2},
          eid = {155},
        pages = {155},
          doi = {10.3847/1538-4357/ad1bcb},
archivePrefix = {arXiv},
       eprint = {2401.03019},
 primaryClass = {astro-ph.HE},
       adsurl = {https://ui.adsabs.harvard.edu/abs/2024ApJ...962..155M}
}

@ARTICLE{2022MNRAS.516.4397K,
       author = {{Kynoch}, Daniel and {Landt}, Hermine and {Dehghanian}, Maryam and {Ward}, Martin J. and {Ferland}, Gary J.},
        title = "{Multiple locations of near-infrared coronal lines in NGC 5548}",
      journal = {\mnras},
         year = 2022,
        month = nov,
       volume = {516},
       number = {3},
        pages = {4397-4416},
          doi = {10.1093/mnras/stac2443},
archivePrefix = {arXiv},
       eprint = {2208.12821},
 primaryClass = {astro-ph.GA},
       adsurl = {https://ui.adsabs.harvard.edu/abs/2022MNRAS.516.4397K}
}

@ARTICLE{2000A&A...354L..83K,
       author = {{Kaastra}, J.~S. and {Mewe}, R. and {Liedahl}, D.~A. and {Komossa}, S. and {Brinkman}, A.~C.},
        title = "{X-ray absorption lines in the Seyfert 1 galaxy NGC 5548 discovered with Chandra-LETGS}",
      journal = {\aap},
         year = 2000,
        month = feb,
       volume = {354},
        pages = {L83-L86},
          doi = {10.48550/arXiv.astro-ph/0002345},
archivePrefix = {arXiv},
       eprint = {astro-ph/0002345},
 primaryClass = {astro-ph},
       adsurl = {https://ui.adsabs.harvard.edu/abs/2000A&A...354L..83K}
}

@ARTICLE{2020ApJ...902...74W,
       author = {{Williams}, P.~R. and {Pancoast}, A. and {Treu}, T. and {Brewer}, B.~J. and {Peterson}, B.~M. and {Barth}, A.~J. and {Malkan}, M.~A. and {De Rosa}, G. and {Horne}, Keith and {Kriss}, G.~A. and {Arav}, N. and {Bentz}, M.~C. and {Cackett}, E.~M. and {Dalla Bont{\`a}}, E. and {Dehghanian}, M. and {Done}, C. and {Ferland}, G.~J. and {Grier}, C.~J. and {Kaastra}, J. and {Kara}, E. and {Kochanek}, C.~S. and {Mathur}, S. and {Mehdipour}, M. and {Pogge}, R.~W. and {Proga}, D. and {Vestergaard}, M. and {Waters}, T. and {Adams}, S.~M. and {Anderson}, M.~D. and {Ar{\'e}valo}, P. and {Beatty}, T.~G. and {Bennert}, V.~N. and {Bigley}, A. and {Bisogni}, S. and {Borman}, G.~A. and {Boroson}, T.~A. and {Bottorff}, M.~C. and {Brandt}, W.~N. and {Breeveld}, A.~A. and {Brotherton}, M. and {Brown}, J.~E. and {Brown}, J.~S. and {Canalizo}, G. and {Carini}, M.~T. and {Clubb}, K.~I. and {Comerford}, J.~M. and {Corsini}, E.~M. and {Crenshaw}, D.~M. and {Croft}, S. and {Croxall}, K.~V. and {Deason}, A.~J. and {De Lorenzo-C{\'a}ceres}, A. and {Denney}, K.~D. and {Dietrich}, M. and {Edelson}, R. and {Efimova}, N.~V. and {Ely}, J. and {Evans}, P.~A. and {Fausnaugh}, M.~M. and {Filippenko}, A.~V. and {Flatland}, K. and {Fox}, O.~D. and {Gardner}, E. and {Gates}, E.~L. and {Gehrels}, N. and {Geier}, S. and {Gelbord}, J.~M. and {Gonzalez}, L. and {Gorjian}, V. and {Greene}, J.~E. and {Grupe}, D. and {Gupta}, A. and {Hall}, P.~B. and {Henderson}, C.~B. and {Hicks}, S. and {Holmbeck}, E. and {Holoien}, T.~W. -S. and {Hutchison}, T. and {Im}, M. and {Jensen}, J.~J. and {Johnson}, C.~A. and {Joner}, M.~D. and {Jones}, J. and {Kaspi}, S. and {Kelly}, P.~L. and {Kennea}, J.~A. and {Kim}, M. and {Kim}, S. and {Kim}, S.~C. and {King}, A. and {Klimanov}, S.~A. and {Knigge}, C. and {Krongold}, Y. and {Lau}, M.~W. and {Lee}, J.~C. and {Leonard}, D.~C. and {Li}, Miao and {Lira}, P. and {Lochhaas}, C. and {Ma}, Zhiyuan and {MacInnis}, F. and {Manne-Nicholas}, E.~R. and {Mauerhan}, J.~C. and {McGurk}, R. and {McHardy}, I.~M. and {Montuori}, C. and {Morelli}, L. and {Mosquera}, A. and {Mudd}, D. and {M{\"u}ller-S{\'a}nchez}, F. and {Nazarov}, S.~V. and {Norris}, R.~P. and {Nousek}, J.~A. and {Nguyen}, M.~L. and {Ochner}, P. and {Okhmat}, D.~N. and {Papadakis}, I. and {Parks}, J.~R. and {Pei}, L. and {Penny}, M.~T. and {Pizzella}, A. and {Poleski}, R. and {Pott}, J. -U. and {Rafter}, S.~E. and {Rix}, H. -W. and {Runnoe}, J. and {Saylor}, D.~A. and {Schimoia}, J.~S. and {Scott}, B. and {Sergeev}, S.~G. and {Shappee}, B.~J. and {Shivvers}, I. and {Siegel}, M. and {Simonian}, G.~V. and {Siviero}, A. and {Skielboe}, A. and {Somers}, G. and {Spencer}, M. and {Starkey}, D. and {Stevens}, D.~J. and {Sung}, H. -I. and {Tayar}, J. and {Tejos}, N. and {Turner}, C.~S. and {Uttley}, P. and {Van Saders}, J. and {Vaughan}, S.~A. and {Vican}, L. and {Villanueva}, Jr., S. and {Villforth}, C. and {Weiss}, Y. and {Woo}, J. -H. and {Yan}, H. and {Young}, S. and {Yuk}, H. and {Zheng}, W. and {Zhu}, W. and {Zu}, Y.},
        title = "{Space Telescope and Optical Reverberation Mapping Project. XII. Broad-line Region Modeling of NGC 5548}",
      journal = {\apj},
         year = 2020,
        month = oct,
       volume = {902},
       number = {1},
          eid = {74},
        pages = {74},
          doi = {10.3847/1538-4357/abbad7},
archivePrefix = {arXiv},
       eprint = {2010.00594},
 primaryClass = {astro-ph.GA},
       adsurl = {https://ui.adsabs.harvard.edu/abs/2020ApJ...902...74W}
}

@ARTICLE{2017MNRAS.464.1771S,
       author = {{Sch{\"o}nell}, Jr., Astor J. and {Storchi-Bergmann}, Thaisa and {Riffel}, Rogemar A. and {Riffel}, Rog{\'e}rio},
        title = "{Feeding versus feedback in active galactic nuclei from near-infrared integral field spectroscopy - XII. NGC 5548}",
      journal = {\mnras},
         year = 2017,
        month = jan,
       volume = {464},
       number = {2},
        pages = {1771-1782},
          doi = {10.1093/mnras/stw2263},
       adsurl = {https://ui.adsabs.harvard.edu/abs/2017MNRAS.464.1771S}
}

@ARTICLE{2016A&A...592A..27C,
       author = {{Cappi}, M. and {De Marco}, B. and {Ponti}, G. and {Ursini}, F. and {Petrucci}, P. -O. and {Bianchi}, S. and {Kaastra}, J.~S. and {Kriss}, G.~A. and {Mehdipour}, M. and {Whewell}, M. and {Arav}, N. and {Behar}, E. and {Boissay}, R. and {Branduardi-Raymont}, G. and {Costantini}, E. and {Ebrero}, J. and {Di Gesu}, L. and {Harrison}, F.~A. and {Kaspi}, S. and {Matt}, G. and {Paltani}, S. and {Peterson}, B.~M. and {Steenbrugge}, K.~C. and {Walton}, D.~J.},
        title = "{Anatomy of the AGN in NGC 5548. VIII. XMM-Newton's EPIC detailed view of an unexpected variable multilayer absorber}",
      journal = {\aap},
         year = 2016,
        month = jul,
       volume = {592},
          eid = {A27},
        pages = {A27},
          doi = {10.1051/0004-6361/201628464},
archivePrefix = {arXiv},
       eprint = {1604.01777},
 primaryClass = {astro-ph.HE},
       adsurl = {https://ui.adsabs.harvard.edu/abs/2016A&A...592A..27C}
}

@ARTICLE{AGN_STORM_2021,
       author = {{Horne}, Keith and {De Rosa}, G. and {Peterson}, B.~M. and {Barth}, A.~J. and {Ely}, J. and {Fausnaugh}, M.~M. and {Kriss}, G.~A. and {Pei}, L. and {Bentz}, M.~C. and {Cackett}, E.~M. and {Edelson}, R. and {Eracleous}, M. and {Goad}, M.~R. and {Grier}, C.~J. and {Kaastra}, J. and {Kochanek}, C.~S. and {Krongold}, Y. and {Mathur}, S. and {Netzer}, H. and {Proga}, D. and {Tejos}, N. and {Vestergaard}, M. and {Villforth}, C. and {Adams}, S.~M. and {Anderson}, M.~D. and {Ar{\'e}valo}, P. and {Beatty}, T.~G. and {Bennert}, V.~N. and {Bigley}, A. and {Bisogni}, S. and {Borman}, G.~A. and {Boroson}, T.~A. and {Bottorff}, M.~C. and {Brandt}, W.~N. and {Breeveld}, A.~A. and {Brotherton}, M. and {Brown}, J.~E. and {Brown}, J.~S. and {Canalizo}, G. and {Carini}, M.~T. and {Clubb}, K.~I. and {Comerford}, J.~M. and {Corsini}, E.~M. and {Crenshaw}, D.~M. and {Croft}, S. and {Croxall}, K.~V. and {Dalla Bont{\`a}}, E. and {Deason}, A.~J. and {Dehghanian}, M. and {De Lorenzo-C{\'a}ceres}, A. and {Denney}, K.~D. and {Dietrich}, M. and {Done}, C. and {Efimova}, N.~V. and {Evans}, P.~A. and {Ferland}, G.~J. and {Filippenko}, A.~V. and {Flatland}, K. and {Fox}, O.~D. and {Gardner}, E. and {Gates}, E.~L. and {Gehrels}, N. and {Geier}, S. and {Gelbord}, J.~M. and {Gonzalez}, L. and {Gorjian}, V. and {Greene}, J.~E. and {Grupe}, D. and {Gupta}, A. and {Hall}, P.~B. and {Henderson}, C.~B. and {Hicks}, S. and {Holmbeck}, E. and {Holoien}, T.~W. -S. and {Hutchison}, T. and {Im}, M. and {Jensen}, J.~J. and {Johnson}, C.~A. and {Joner}, M.~D. and {Jones}, J. and {Kaspi}, S. and {Kelly}, P.~L. and {Kennea}, J.~A. and {Kim}, M. and {Kim}, S. and {Kim}, S.~C. and {King}, A. and {Klimanov}, S.~A. and {Korista}, K.~T. and {Lau}, M.~W. and {Lee}, J.~C. and {Leonard}, D.~C. and {Li}, Miao and {Lira}, P. and {Lochhaas}, C. and {Ma}, Zhiyuan and {MacInnis}, F. and {Malkan}, M.~A. and {Manne-Nicholas}, E.~R. and {Mauerhan}, J.~C. and {McGurk}, R. and {McHardy}, I.~M. and {Montuori}, C. and {Morelli}, L. and {Mosquera}, A. and {Mudd}, D. and {M{\"u}ller-S{\'a}nchez}, F. and {Nazarov}, S.~V. and {Norris}, R.~P. and {Nousek}, J.~A. and {Nguyen}, M.~L. and {Ochner}, P. and {Okhmat}, D.~N. and {Pancoast}, A. and {Papadakis}, I. and {Parks}, J.~R. and {Penny}, M.~T. and {Pizzella}, A. and {Pogge}, R.~W. and {Poleski}, R. and {Pott}, J. -U. and {Rafter}, S.~E. and {Rix}, H. -W. and {Runnoe}, J. and {Saylor}, D.~A. and {Schimoia}, J.~S. and {Schn{\"u}lle}, K. and {Scott}, B. and {Sergeev}, S.~G. and {Shappee}, B.~J. and {Shivvers}, I. and {Siegel}, M. and {Simonian}, G.~V. and {Siviero}, A. and {Skielboe}, A. and {Somers}, G. and {Spencer}, M. and {Starkey}, D. and {Stevens}, D.~J. and {Sung}, H. -I. and {Tayar}, J. and {Treu}, T. and {Turner}, C.~S. and {Uttley}, P. and {Van Saders}, J. and {Vican}, L. and {Villanueva}, Jr., S. and {Weiss}, Y. and {Woo}, J. -H. and {Yan}, H. and {Young}, S. and {Yuk}, H. and {Zheng}, W. and {Zhu}, W. and {Zu}, Y.},
        title = "{Space Telescope and Optical Reverberation Mapping Project. IX. Velocity-Delay Maps for Broad Emission Lines in NGC 5548}",
      journal = {\apj},
         year = 2021,
        month = feb,
       volume = {907},
       number = {2},
          eid = {76},
        pages = {76},
          doi = {10.3847/1538-4357/abce60},
archivePrefix = {arXiv},
       eprint = {2003.01448},
 primaryClass = {astro-ph.GA},
       adsurl = {https://ui.adsabs.harvard.edu/abs/2021ApJ...907...76H}
}

@ARTICLE{Kaastra_2014_ngc5548_2013_campaign,
       author = {{Kaastra}, J.~S. and {Kriss}, G.~A. and {Cappi}, M. and {Mehdipour}, M. and {Petrucci}, P. -O. and {Steenbrugge}, K.~C. and {Arav}, N. and {Behar}, E. and {Bianchi}, S. and {Boissay}, R. and {Branduardi-Raymont}, G. and {Chamberlain}, C. and {Costantini}, E. and {Ely}, J.~C. and {Ebrero}, J. and {Di Gesu}, L. and {Harrison}, F.~A. and {Kaspi}, S. and {Malzac}, J. and {De Marco}, B. and {Matt}, G. and {Nandra}, K. and {Paltani}, S. and {Person}, R. and {Peterson}, B.~M. and {Pinto}, C. and {Ponti}, G. and {Pozo Nu{\~n}ez}, F. and {De Rosa}, A. and {Seta}, H. and {Ursini}, F. and {de Vries}, C.~P. and {Walton}, D.~J. and {Whewell}, M.},
        title = "{A fast and long-lived outflow from the supermassive black hole in NGC 5548}",
      journal = {Science},
         year = 2014,
        month = jul,
       volume = {345},
       number = {6192},
        pages = {64-68},
          doi = {10.1126/science.1253787},
archivePrefix = {arXiv},
       eprint = {1406.5007},
 primaryClass = {astro-ph.HE},
       adsurl = {https://ui.adsabs.harvard.edu/abs/2014Sci...345...64K}
}

@ARTICLE{Mehdipour_2015_ngc5548_2013_campaign,
       author = {{Mehdipour}, M. and {Kaastra}, J.~S. and {Kriss}, G.~A. and {Cappi}, M. and {Petrucci}, P. -O. and {Steenbrugge}, K.~C. and {Arav}, N. and {Behar}, E. and {Bianchi}, S. and {Boissay}, R. and {Branduardi-Raymont}, G. and {Costantini}, E. and {Ebrero}, J. and {Di Gesu}, L. and {Harrison}, F.~A. and {Kaspi}, S. and {De Marco}, B. and {Matt}, G. and {Paltani}, S. and {Peterson}, B.~M. and {Ponti}, G. and {Pozo Nu{\~n}ez}, F. and {De Rosa}, A. and {Ursini}, F. and {de Vries}, C.~P. and {Walton}, D.~J. and {Whewell}, M.},
        title = "{Anatomy of the AGN in NGC 5548. I. A global model for the broadband spectral energy distribution}",
      journal = {\aap},
         year = 2015,
        month = mar,
       volume = {575},
          eid = {A22},
        pages = {A22},
          doi = {10.1051/0004-6361/201425373},
archivePrefix = {arXiv},
       eprint = {1501.01188},
 primaryClass = {astro-ph.HE},
       adsurl = {https://ui.adsabs.harvard.edu/abs/2015A&A...575A..22M}
}

@ARTICLE{AGN_STORM,
       author = {{De Rosa}, G. and {Peterson}, B.~M. and {Ely}, J. and {Kriss}, G.~A. and {Crenshaw}, D.~M. and {Horne}, Keith and {Korista}, K.~T. and {Netzer}, H. and {Pogge}, R.~W. and {Ar{\'e}valo}, P. and {Barth}, A.~J. and {Bentz}, M.~C. and {Brandt}, W.~N. and {Breeveld}, A.~A. and {Brewer}, B.~J. and {Dalla Bont{\`a}}, E. and {De Lorenzo-C{\'a}ceres}, A. and {Denney}, K.~D. and {Dietrich}, M. and {Edelson}, R. and {Evans}, P.~A. and {Fausnaugh}, M.~M. and {Gehrels}, N. and {Gelbord}, J.~M. and {Goad}, M.~R. and {Grier}, C.~J. and {Grupe}, D. and {Hall}, P.~B. and {Kaastra}, J. and {Kelly}, B.~C. and {Kennea}, J.~A. and {Kochanek}, C.~S. and {Lira}, P. and {Mathur}, S. and {McHardy}, I.~M. and {Nousek}, J.~A. and {Pancoast}, A. and {Papadakis}, I. and {Pei}, L. and {Schimoia}, J.~S. and {Siegel}, M. and {Starkey}, D. and {Treu}, T. and {Uttley}, P. and {Vaughan}, S. and {Vestergaard}, M. and {Villforth}, C. and {Yan}, H. and {Young}, S. and {Zu}, Y.},
        title = "{Space Telescope and Optical Reverberation Mapping Project.I. Ultraviolet Observations of the Seyfert 1 Galaxy NGC 5548 with the Cosmic Origins Spectrograph on Hubble Space Telescope}",
      journal = {\apj},
         year = 2015,
        month = jun,
       volume = {806},
       number = {1},
          eid = {128},
        pages = {128},
          doi = {10.1088/0004-637X/806/1/128},
archivePrefix = {arXiv},
       eprint = {1501.05954},
 primaryClass = {astro-ph.GA},
       adsurl = {https://ui.adsabs.harvard.edu/abs/2015ApJ...806..128D}
}

@ARTICLE{DRW_2009,
       author = {{Kelly}, Brandon C. and {Bechtold}, Jill and {Siemiginowska}, Aneta},
        title = "{Are the Variations in Quasar Optical Flux Driven by Thermal Fluctuations?}",
      journal = {\apj},
         year = 2009,
        month = jun,
       volume = {698},
       number = {1},
        pages = {895-910},
          doi = {10.1088/0004-637X/698/1/895},
archivePrefix = {arXiv},
       eprint = {0903.5315},
 primaryClass = {astro-ph.CO},
       adsurl = {https://ui.adsabs.harvard.edu/abs/2009ApJ...698..895K}
}

@ARTICLE{DRW_2010,
       author = {{Koz{\l}owski}, Szymon and {Kochanek}, Christopher S. and {Udalski}, A. and {Wyrzykowski}, {\L}. and {Soszy{\'n}ski}, I. and {Szyma{\'n}ski}, M.~K. and {Kubiak}, M. and {Pietrzy{\'n}ski}, G. and {Szewczyk}, O. and {Ulaczyk}, K. and {Poleski}, R. and {OGLE Collaboration}},
        title = "{Quantifying Quasar Variability as Part of a General Approach to Classifying Continuously Varying Sources}",
      journal = {\apj},
         year = 2010,
        month = jan,
       volume = {708},
       number = {2},
        pages = {927-945},
          doi = {10.1088/0004-637X/708/2/927},
archivePrefix = {arXiv},
       eprint = {0909.1326},
 primaryClass = {astro-ph.CO},
       adsurl = {https://ui.adsabs.harvard.edu/abs/2010ApJ...708..927K}
}

@ARTICLE{2016ApJ...824...11G,
       author = {{Goad}, M.~R. and {Korista}, K.~T. and {De Rosa}, G. and {Kriss}, G.~A. and {Edelson}, R. and {Barth}, A.~J. and {Ferland}, G.~J. and {Kochanek}, C.~S. and {Netzer}, H. and {Peterson}, B.~M. and {Bentz}, M.~C. and {Bisogni}, S. and {Crenshaw}, D.~M. and {Denney}, K.~D. and {Ely}, J. and {Fausnaugh}, M.~M. and {Grier}, C.~J. and {Gupta}, A. and {Horne}, K.~D. and {Kaastra}, J. and {Pancoast}, A. and {Pei}, L. and {Pogge}, R.~W. and {Skielboe}, A. and {Starkey}, D. and {Vestergaard}, M. and {Zu}, Y. and {Anderson}, M.~D. and {Ar{\'e}valo}, P. and {Bazhaw}, C. and {Borman}, G.~A. and {Boroson}, T.~A. and {Bottorff}, M.~C. and {Brandt}, W.~N. and {Breeveld}, A.~A. and {Brewer}, B.~J. and {Cackett}, E.~M. and {Carini}, M.~T. and {Croxall}, K.~V. and {Dalla Bont{\`a}}, E. and {De Lorenzo-C{\'a}ceres}, A. and {Dietrich}, M. and {Efimova}, N.~V. and {Evans}, P.~A. and {Filippenko}, A.~V. and {Flatland}, K. and {Gehrels}, N. and {Geier}, S. and {Gelbord}, J.~M. and {Gonzalez}, L. and {Gorjian}, V. and {Grupe}, D. and {Hall}, P.~B. and {Hicks}, S. and {Horenstein}, D. and {Hutchison}, T. and {Im}, M. and {Jensen}, J.~J. and {Joner}, M.~D. and {Jones}, J. and {Kaspi}, S. and {Kelly}, B.~C. and {Kennea}, J.~A. and {Kim}, M. and {Kim}, S.~C. and {Klimanov}, S.~A. and {Lee}, J.~C. and {Leonard}, D.~C. and {Lira}, P. and {MacInnis}, F. and {Manne-Nicholas}, E.~R. and {Mathur}, S. and {McHardy}, I.~M. and {Montouri}, C. and {Musso}, R. and {Nazarov}, S.~V. and {Norris}, R.~P. and {Nousek}, J.~A. and {Okhmat}, D.~N. and {Papadakis}, I. and {Parks}, J.~R. and {Pott}, J. -U. and {Rafter}, S.~E. and {Rix}, H. -W. and {Saylor}, D.~A. and {Schimoia}, J.~S. and {Schn{\"u}lle}, K. and {Sergeev}, S.~G. and {Siegel}, M. and {Spencer}, M. and {Sung}, H. -I. and {Teems}, K.~G. and {Treu}, T. and {Turner}, C.~S. and {Uttley}, P. and {Villforth}, C. and {Weiss}, Y. and {Woo}, J. -H. and {Yan}, H. and {Young}, S. and {Zheng}, W. -K.},
        title = "{Space Telescope and Optical Reverberation Mapping Project. IV. Anomalous Behavior of the Broad Ultraviolet Emission Lines in NGC 5548}",
      journal = {\apj},
         year = 2016,
        month = jun,
       volume = {824},
       number = {1},
          eid = {11},
        pages = {11},
          doi = {10.3847/0004-637X/824/1/11},
archivePrefix = {arXiv},
       eprint = {1603.08741},
 primaryClass = {astro-ph.GA},
       adsurl = {https://ui.adsabs.harvard.edu/abs/2016ApJ...824...11G}
}

@ARTICLE{2017ApJ...837..131P,
       author = {{Pei}, L. and {Fausnaugh}, M.~M. and {Barth}, A.~J. and {Peterson}, B.~M. and {Bentz}, M.~C. and {De Rosa}, G. and {Denney}, K.~D. and {Goad}, M.~R. and {Kochanek}, C.~S. and {Korista}, K.~T. and {Kriss}, G.~A. and {Pogge}, R.~W. and {Bennert}, V.~N. and {Brotherton}, M. and {Clubb}, K.~I. and {Dalla Bont{\`a}}, E. and {Filippenko}, A.~V. and {Greene}, J.~E. and {Grier}, C.~J. and {Vestergaard}, M. and {Zheng}, W. and {Adams}, Scott M. and {Beatty}, Thomas G. and {Bigley}, A. and {Brown}, Jacob E. and {Brown}, Jonathan S. and {Canalizo}, G. and {Comerford}, J.~M. and {Coker}, Carl T. and {Corsini}, E.~M. and {Croft}, S. and {Croxall}, K.~V. and {Deason}, A.~J. and {Eracleous}, Michael and {Fox}, O.~D. and {Gates}, E.~L. and {Henderson}, C.~B. and {Holmbeck}, E. and {Holoien}, T.~W. -S. and {Jensen}, J.~J. and {Johnson}, C.~A. and {Kelly}, P.~L. and {Kim}, S. and {King}, A. and {Lau}, M.~W. and {Li}, Miao and {Lochhaas}, Cassandra and {Ma}, Zhiyuan and {Manne-Nicholas}, E.~R. and {Mauerhan}, J.~C. and {Malkan}, M.~A. and {McGurk}, R. and {Morelli}, L. and {Mosquera}, Ana and {Mudd}, Dale and {Muller Sanchez}, F. and {Nguyen}, M.~L. and {Ochner}, P. and {Ou-Yang}, B. and {Pancoast}, A. and {Penny}, Matthew T. and {Pizzella}, A. and {Poleski}, Rados{\l}aw and {Runnoe}, Jessie and {Scott}, B. and {Schimoia}, Jaderson S. and {Shappee}, B.~J. and {Shivvers}, I. and {Simonian}, Gregory V. and {Siviero}, A. and {Somers}, Garrett and {Stevens}, Daniel J. and {Strauss}, M.~A. and {Tayar}, Jamie and {Tejos}, N. and {Treu}, T. and {Van Saders}, J. and {Vican}, L. and {Villanueva}, Jr., S. and {Yuk}, H. and {Zakamska}, N.~L. and {Zhu}, W. and {Anderson}, M.~D. and {Ar{\'e}valo}, P. and {Bazhaw}, C. and {Bisogni}, S. and {Borman}, G.~A. and {Bottorff}, M.~C. and {Brandt}, W.~N. and {Breeveld}, A.~A. and {Cackett}, E.~M. and {Carini}, M.~T. and {Crenshaw}, D.~M. and {De Lorenzo-C{\'a}ceres}, A. and {Dietrich}, M. and {Edelson}, R. and {Efimova}, N.~V. and {Ely}, J. and {Evans}, P.~A. and {Ferland}, G.~J. and {Flatland}, K. and {Gehrels}, N. and {Geier}, S. and {Gelbord}, J.~M. and {Grupe}, D. and {Gupta}, A. and {Hall}, P.~B. and {Hicks}, S. and {Horenstein}, D. and {Horne}, Keith and {Hutchison}, T. and {Im}, M. and {Joner}, M.~D. and {Jones}, J. and {Kaastra}, J. and {Kaspi}, S. and {Kelly}, B.~C. and {Kennea}, J.~A. and {Kim}, M. and {Kim}, S.~C. and {Klimanov}, S.~A. and {Lee}, J.~C. and {Leonard}, D.~C. and {Lira}, P. and {MacInnis}, F. and {Mathur}, S. and {McHardy}, I.~M. and {Montouri}, C. and {Musso}, R. and {Nazarov}, S.~V. and {Netzer}, H. and {Norris}, R.~P. and {Nousek}, J.~A. and {Okhmat}, D.~N. and {Papadakis}, I. and {Parks}, J.~R. and {Pott}, J. -U. and {Rafter}, S.~E. and {Rix}, H. -W. and {Saylor}, D.~A. and {Schn{\"u}lle}, K. and {Sergeev}, S.~G. and {Siegel}, M. and {Skielboe}, A. and {Spencer}, M. and {Starkey}, D. and {Sung}, H. -I. and {Teems}, K.~G. and {Turner}, C.~S. and {Uttley}, P. and {Villforth}, C. and {Weiss}, Y. and {Woo}, J. -H. and {Yan}, H. and {Young}, S. and {Zu}, Y.},
        title = "{Space Telescope and Optical Reverberation Mapping Project. V. Optical Spectroscopic Campaign and Emission-line Analysis for NGC 5548}",
      journal = {\apj},
         year = 2017,
        month = mar,
       volume = {837},
       number = {2},
          eid = {131},
        pages = {131},
          doi = {10.3847/1538-4357/aa5eb1},
archivePrefix = {arXiv},
       eprint = {1702.01177},
 primaryClass = {astro-ph.GA},
       adsurl = {https://ui.adsabs.harvard.edu/abs/2017ApJ...837..131P}
}

@ARTICLE{2019MNRAS.486.5362G,
       author = {{Goad}, M.~R. and {Knigge}, C. and {Korista}, K.~T. and {Cackett}, E. and {Horne}, K. and {Starkey}, D.~A. and {Peterson}, B.~M. and {De Rosa}, G. and {Kriss}, G.~A. and {Edelson}, R. and {Fausnaugh}, M.},
        title = "{Anomalous behaviour of the UV-optical continuum bands in NGC 5548}",
      journal = {\mnras},
         year = 2019,
        month = jul,
       volume = {486},
       number = {4},
        pages = {5362-5376},
          doi = {10.1093/mnras/stz1186},
archivePrefix = {arXiv},
       eprint = {1904.12588},
 primaryClass = {astro-ph.GA},
       adsurl = {https://ui.adsabs.harvard.edu/abs/2019MNRAS.486.5362G}
}

@ARTICLE{Cloudy2023,
       author = {{Chatzikos}, M. and {Bianchi}, S. and {Camilloni}, F. and {Chakraborty}, P. and {Gunasekera}, C.~M. and {Guzm{\'a}n}, F. and {Milby}, J.~S. and {Sarkar}, A. and {Shaw}, G. and {van Hoof}, P.~A.~M. and {Ferland}, G.~J.},
        title = "{The 2023 Release of Cloudy}",
      journal = {\rmxaa},
         year = 2023,
        month = oct,
       volume = {59},
        pages = {327-343},
          doi = {10.22201/ia.01851101p.2023.59.02.12},
archivePrefix = {arXiv},
       eprint = {2308.06396},
 primaryClass = {astro-ph.GA},
       adsurl = {https://ui.adsabs.harvard.edu/abs/2023RMxAA..59..327C}
}

@ARTICLE{CHIANTI,
       author = {{Del Zanna}, G. and {Dere}, K.~P. and {Young}, P.~R. and {Landi}, E. and {Mason}, H.~E.},
        title = "{CHIANTI - An atomic database for emission lines. Version 8}",
      journal = {\aap},
         year = 2015,
        month = oct,
       volume = {582},
          eid = {A56},
        pages = {A56},
          doi = {10.1051/0004-6361/201526827},
archivePrefix = {arXiv},
       eprint = {1508.07631},
 primaryClass = {astro-ph.SR},
       adsurl = {https://ui.adsabs.harvard.edu/abs/2015A&A...582A..56D}
}

@ARTICLE{MacLeod_2010_DRW,
       author = {{MacLeod}, C.~L. and {Ivezi{\'c}}, {\v{Z}}. and {Kochanek}, C.~S. and {Koz{\l}owski}, S. and {Kelly}, B. and {Bullock}, E. and {Kimball}, A. and {Sesar}, B. and {Westman}, D. and {Brooks}, K. and {Gibson}, R. and {Becker}, A.~C. and {de Vries}, W.~H.},
        title = "{Modeling the Time Variability of SDSS Stripe 82 Quasars as a Damped Random Walk}",
      journal = {\apj},
         year = 2010,
        month = oct,
       volume = {721},
       number = {2},
        pages = {1014-1033},
          doi = {10.1088/0004-637X/721/2/1014},
archivePrefix = {arXiv},
       eprint = {1004.0276},
 primaryClass = {astro-ph.CO},
       adsurl = {https://ui.adsabs.harvard.edu/abs/2010ApJ...721.1014M}
}

@ARTICLE{MacLeod_2012_DRW,
       author = {{MacLeod}, Chelsea L. and {Ivezi{\'c}}, {\v{Z}}eljko and {Sesar}, Branimir and {de Vries}, Wim and {Kochanek}, Christopher S. and {Kelly}, Brandon C. and {Becker}, Andrew C. and {Lupton}, Robert H. and {Hall}, Patrick B. and {Richards}, Gordon T. and {Anderson}, Scott F. and {Schneider}, Donald P.},
        title = "{A Description of Quasar Variability Measured Using Repeated SDSS and POSS Imaging}",
      journal = {\apj},
         year = 2012,
        month = jul,
       volume = {753},
       number = {2},
          eid = {106},
        pages = {106},
          doi = {10.1088/0004-637X/753/2/106},
archivePrefix = {arXiv},
       eprint = {1112.0679},
 primaryClass = {astro-ph.CO},
       adsurl = {https://ui.adsabs.harvard.edu/abs/2012ApJ...753..106M}
}

@ARTICLE{Yu_2020,
       author = {{Yu}, Z. and {Kochanek}, C.~S. and {Peterson}, B.~M. and {Zu}, Y. and {Brandt}, W.~N. and {Cackett}, E.~M. and {Fausnaugh}, M.~M. and {McHardy}, I.~M.},
        title = "{On reverberation mapping lag uncertainties}",
      journal = {\mnras},
         year = 2020,
        month = feb,
       volume = {491},
       number = {4},
        pages = {6045-6064},
          doi = {10.1093/mnras/stz3464},
archivePrefix = {arXiv},
       eprint = {1909.03072},
 primaryClass = {astro-ph.GA},
       adsurl = {https://ui.adsabs.harvard.edu/abs/2020MNRAS.491.6045Y}
}

@ARTICLE{2017ApJ...847..132G,
       author = {{Guo}, Hengxiao and {Wang}, Junxian and {Cai}, Zhenyi and {Sun}, Mouyuan},
        title = "{How Far Is Quasar UV/Optical Variability from a Damped Random Walk at Low Frequency?}",
      journal = {\apj},
         year = 2017,
        month = oct,
       volume = {847},
       number = {2},
          eid = {132},
        pages = {132},
          doi = {10.3847/1538-4357/aa8d71},
archivePrefix = {arXiv},
       eprint = {1709.05271},
 primaryClass = {astro-ph.GA},
       adsurl = {https://ui.adsabs.harvard.edu/abs/2017ApJ...847..132G}
}

@ARTICLE{Crenshaw_2003,
       author = {{Crenshaw}, D.~M. and {Kraemer}, S.~B. and {Gabel}, J.~R. and {Kaastra}, J.~S. and {Steenbrugge}, K.~C. and {Brinkman}, A.~C. and {Dunn}, J.~P. and {George}, I.~M. and {Liedahl}, D.~A. and {Paerels}, F.~B.~S. and {Turner}, T.~J. and {Yaqoob}, T.},
        title = "{Simultaneous Ultraviolet and X-Ray Spectroscopy of the Seyfert 1 Galaxy NGC 5548. I. Physical Conditions in the Ultraviolet Absorbers}",
      journal = {\apj},
         year = 2003,
        month = sep,
       volume = {594},
       number = {1},
        pages = {116-127},
          doi = {10.1086/376792},
archivePrefix = {arXiv},
       eprint = {astro-ph/0305154},
 primaryClass = {astro-ph},
       adsurl = {https://ui.adsabs.harvard.edu/abs/2003ApJ...594..116C}
}

@ARTICLE{D22,
       author = {{Dov{\v{c}}iak}, M. and {Papadakis}, I.~E. and {Kammoun}, E.~S. and {Zhang}, W.},
        title = "{Physical model for the broadband energy spectrum of X-ray illuminated accretion discs: Fitting the spectral energy distribution of NGC 5548}",
      journal = {\aap},
         year = 2022,
        month = may,
       volume = {661},
          eid = {A135},
        pages = {A135},
          doi = {10.1051/0004-6361/202142358},
archivePrefix = {arXiv},
       eprint = {2110.01249},
 primaryClass = {astro-ph.HE},
       adsurl = {https://ui.adsabs.harvard.edu/abs/2022A&A...661A.135D}
}

@ARTICLE{Filiz_Ak_2013,
       author = {{Filiz Ak}, N. and {Brandt}, W.~N. and {Hall}, P.~B. and {Schneider}, D.~P. and {Anderson}, S.~F. and {Hamann}, F. and {Lundgren}, B.~F. and {Myers}, Adam D. and {P{\^a}ris}, I. and {Petitjean}, P. and {Ross}, Nicholas P. and {Shen}, Yue and {York}, Don},
        title = "{Broad Absorption Line Variability on Multi-year Timescales in a Large Quasar Sample}",
      journal = {\apj},
         year = 2013,
        month = nov,
       volume = {777},
       number = {2},
          eid = {168},
        pages = {168},
          doi = {10.1088/0004-637X/777/2/168},
archivePrefix = {arXiv},
       eprint = {1309.5364},
 primaryClass = {astro-ph.CO},
       adsurl = {https://ui.adsabs.harvard.edu/abs/2013ApJ...777..168F}
}

@ARTICLE{2019ApJ...881..153K,
       author = {{Kriss}, G.~A. and {De Rosa}, G. and {Ely}, J. and {Peterson}, B.~M. and {Kaastra}, J. and {Mehdipour}, M. and {Ferland}, G.~J. and {Dehghanian}, M. and {Mathur}, S. and {Edelson}, R. and {Korista}, K.~T. and {Arav}, N. and {Barth}, A.~J. and {Bentz}, M.~C. and {Brandt}, W.~N. and {Crenshaw}, D.~M. and {Dalla Bont{\`a}}, E. and {Denney}, K.~D. and {Done}, C. and {Eracleous}, M. and {Fausnaugh}, M.~M. and {Gardner}, E. and {Goad}, M.~R. and {Grier}, C.~J. and {Horne}, Keith and {Kochanek}, C.~S. and {McHardy}, I.~M. and {Netzer}, H. and {Pancoast}, A. and {Pei}, L. and {Pogge}, R.~W. and {Proga}, D. and {Silva}, C. and {Tejos}, N. and {Vestergaard}, M. and {Adams}, S.~M. and {Anderson}, M.~D. and {Ar{\'e}valo}, P. and {Beatty}, T.~G. and {Behar}, E. and {Bennert}, V.~N. and {Bianchi}, S. and {Bigley}, A. and {Bisogni}, S. and {Boissay-Malaquin}, R. and {Borman}, G.~A. and {Bottorff}, M.~C. and {Breeveld}, A.~A. and {Brotherton}, M. and {Brown}, J.~E. and {Brown}, J.~S. and {Cackett}, E.~M. and {Canalizo}, G. and {Cappi}, M. and {Carini}, M.~T. and {Clubb}, K.~I. and {Comerford}, J.~M. and {Coker}, C.~T. and {Corsini}, E.~M. and {Costantini}, E. and {Croft}, S. and {Croxall}, K.~V. and {Deason}, A.~J. and {De Lorenzo-C{\'a}ceres}, A. and {De Marco}, B. and {Dietrich}, M. and {Di Gesu}, L. and {Ebrero}, J. and {Evans}, P.~A. and {Filippenko}, A.~V. and {Flatland}, K. and {Gates}, E.~L. and {Gehrels}, N. and {Geier}, S. and {Gelbord}, J.~M. and {Gonzalez}, L. and {Gorjian}, V. and {Grupe}, D. and {Gupta}, A. and {Hall}, P.~B. and {Henderson}, C.~B. and {Hicks}, S. and {Holmbeck}, E. and {Holoien}, T.~W. -S. and {Hutchison}, T.~A. and {Im}, M. and {Jensen}, J.~J. and {Johnson}, C.~A. and {Joner}, M.~D. and {Kaspi}, S. and {Kelly}, B.~C. and {Kelly}, P.~L. and {Kennea}, J.~A. and {Kim}, M. and {Kim}, S.~C. and {Kim}, S.~Y. and {King}, A. and {Klimanov}, S.~A. and {Krongold}, Y. and {Lau}, M.~W. and {Lee}, J.~C. and {Leonard}, D.~C. and {Li}, Miao and {Lira}, P. and {Lochhaas}, C. and {Ma}, Zhiyuan and {MacInnis}, F. and {Malkan}, M.~A. and {Manne-Nicholas}, E.~R. and {Matt}, G. and {Mauerhan}, J.~C. and {McGurk}, R. and {Montuori}, C. and {Morelli}, L. and {Mosquera}, A. and {Mudd}, D. and {M{\"u}ller-S{\'a}nchez}, F. and {Nazarov}, S.~V. and {Norris}, R.~P. and {Nousek}, J.~A. and {Nguyen}, M.~L. and {Ochner}, P. and {Okhmat}, D.~N. and {Paltani}, S. and {Parks}, J.~R. and {Pinto}, C. and {Pizzella}, A. and {Poleski}, R. and {Ponti}, G. and {Pott}, J. -U. and {Rafter}, S.~E. and {Rix}, H. -W. and {Runnoe}, J. and {Saylor}, D.~A. and {Schimoia}, J.~S. and {Schn{\"u}lle}, K. and {Scott}, B. and {Sergeev}, S.~G. and {Shappee}, B.~J. and {Shivvers}, I. and {Siegel}, M. and {Simonian}, G.~V. and {Siviero}, A. and {Skielboe}, A. and {Somers}, G. and {Spencer}, M. and {Starkey}, D. and {Stevens}, D.~J. and {Sung}, H. -I. and {Tayar}, J. and {Teems}, K.~G. and {Treu}, T. and {Turner}, C.~S. and {Uttley}, P. and {. Van Saders}, J. and {Vican}, L. and {Villforth}, C. and {Villanueva}, Jr., S. and {Walton}, D.~J. and {Waters}, T. and {Weiss}, Y. and {Woo}, J. -H. and {Yan}, H. and {Yuk}, H. and {Zheng}, W. and {Zhu}, W. and {Zu}, Y.},
        title = "{Space Telescope and Optical Reverberation Mapping Project. VIII. Time Variability of Emission and Absorption in NGC 5548 Based on Modeling the Ultraviolet Spectrum}",
      journal = {\apj},
         year = 2019,
        month = aug,
       volume = {881},
       number = {2},
          eid = {153},
        pages = {153},
          doi = {10.3847/1538-4357/ab3049},
archivePrefix = {arXiv},
       eprint = {1907.03874},
 primaryClass = {astro-ph.GA},
       adsurl = {https://ui.adsabs.harvard.edu/abs/2019ApJ...881..153K}
}

@article{Chen_2026,
    doi = {10.3847/1538-4357/ae2858},
    url = {https://doi.org/10.3847/1538-4357/ae2858},
    year = {2026},
    month = {jan},
    publisher = {The American Astronomical Society},
    volume = {997},
    number = {2},
    pages = {245},
    author = {Chen, Yaqi and He, Zhicheng and Liu, Guilin},
    title = {Dissecting the Multiple Component Outflow in NGC 5548 with Absorption Line Variability},
    journal = {The Astrophysical Journal}
}

@ARTICLE{2024SCPMA..6729512H,
       author = {{He}, Zhicheng and {Chen}, Zhifu and {Liu}, Guilin and {Wang}, Tinggui and {Ho}, Luis C. and {Wang}, Junxian and {Bian}, Weihao and {Cai}, Zheng and {Mou}, Guobin and {Gu}, Qiusheng and {Wang}, Zhiwen},
        title = "{The transition from galaxy-wide gas inflow to outflow in quasar host galaxies}",
      journal = {Science China Physics, Mechanics, and Astronomy},
         year = 2024,
        month = dec,
       volume = {67},
       number = {12},
          eid = {129512},
        pages = {129512},
          doi = {10.1007/s11433-024-2475-7},
archivePrefix = {arXiv},
       eprint = {2408.04458},
 primaryClass = {astro-ph.GA},
       adsurl = {https://ui.adsabs.harvard.edu/abs/2024SCPMA..6729512H}
}

@ARTICLE{2025A&A...703A.305H,
       author = {{He}, Zhicheng and {Wang}, Tinggui},
        title = "{Discovery of the asymmetric effect in the response of photoionization gas}",
      journal = {\aap},
         year = 2025,
        month = nov,
       volume = {703},
          eid = {A305},
        pages = {A305},
          doi = {10.1051/0004-6361/202556191},
archivePrefix = {arXiv},
       eprint = {2311.16463},
 primaryClass = {astro-ph.GA},
       adsurl = {https://ui.adsabs.harvard.edu/abs/2025A&A...703A.305H}
}

@ARTICLE{2021ApJ...906L...8Z,
       author = {{Zhao}, Qinyuan and {He}, Zhicheng and {Liu}, Guilin and {Wang}, Tinggui and {Guo}, Hengxiao and {Shen}, Lu and {Mou}, Guobin},
        title = "{A Sharp Rise in the Detection Rate of Broad Absorption Line Variations in a Quasar SDSS J141955.26+522741.1}",
      journal = {\apjl},
         year = 2021,
        month = jan,
       volume = {906},
       number = {2},
          eid = {L8},
        pages = {L8},
          doi = {10.3847/2041-8213/abd318},
archivePrefix = {arXiv},
       eprint = {2012.07254},
 primaryClass = {astro-ph.GA},
       adsurl = {https://ui.adsabs.harvard.edu/abs/2021ApJ...906L...8Z}
}

@ARTICLE{2022NatAs...6..339C,
       author = {{Chen}, Zhifu and {He}, Zhicheng and {Ho}, Luis C. and {Gu}, Qiusheng and {Wang}, Tinggui and {Zhuang}, Mingyang and {Liu}, Guilin and {Wang}, Zhiwen},
        title = "{Evidence for the connection between star formation rate and the evolutionary phases of quasars}",
      journal = {Nature Astronomy},
         year = 2022,
        month = jan,
       volume = {6},
        pages = {339-343},
          doi = {10.1038/s41550-021-01561-3},
archivePrefix = {arXiv},
       eprint = {2111.09594},
 primaryClass = {astro-ph.GA},
       adsurl = {https://ui.adsabs.harvard.edu/abs/2022NatAs...6..339C}
}
\bibliographystyle{aasjournalv7}



\end{document}